\documentclass[preprint,12pt]{aastex}

\usepackage{apjfonts}
\usepackage{epsfig}
\usepackage{graphicx}
\usepackage{url}
\usepackage{natbib}
\usepackage{float}

\shorttitle{Sgr Leading Tidal Tail}
\shortauthors{Newby et al.}

\begin{document}

\title{ A Spatial Characterization of the Sagittarius Dwarf Galaxy Tidal Tails }

\author{
Matthew Newby,\altaffilmark{\ref{RPI_A}}
Nathan Cole,\altaffilmark{\ref{RPI_A}}
Heidi Jo Newberg,\altaffilmark{\ref{RPI_A}}
Travis Desell,\altaffilmark{\ref{ND}}
Malik Magdon-Ismail,\altaffilmark{\ref{RPI_CS}}
Boleslaw Szymanski,\altaffilmark{\ref{RPI_CS}}
Carlos Varela,\altaffilmark{\ref{RPI_CS}}
Benjamin Willett,\altaffilmark{\ref{RPI_A}}
Brian Yanny \altaffilmark{\ref{fnal}}
}

\altaffiltext{1}{Department of Physics, Applied Physics and Astronomy, Rensselaer
Polytechnic Institute, Troy, NY 12180, USA; newbym2@rpi.edu, heidi@rpi.edu\label{RPI_A}}


\altaffiltext{2}{Department of Computer Science, U. of North Dakota, Grand Forks, ND 52802, USA\label{ND}}

\altaffiltext{3}{Department of Computer Science, Rensselaer
Polytechnic Institute, Troy, NY 12180, USA\label{RPI_CS}}

\altaffiltext{4}{Fermi National Accelerator Laboratory, P.O. Box 500, Batavia, IL, 60510, USA \label{fnal}}

\begin{abstract}
We measure the spatial density of F turnoff stars in the Sagittarius dwarf tidal stream, from Sloan 
Digital Sky Survey (SDSS) data, using statistical photometric parallax.  We find a set of continuous, 
consistent parameters that describe the leading Sgr stream's position, direction, and width for 15 
stripes in the North Galactic Cap, and 3 stripes in the South Galactic Cap. 
We produce a catalog of stars that has the density characteristics of the dominant leading Sgr tidal 
stream that can be compared with simulations.  We find that the width of the leading (North) tidal 
tail is consistent with recent triaxial and axisymmetric halo model simulations.
The density along the stream is roughly consistent common disruption models in the North, but possibly not in the South.
We explore the possibility that one or more of the dominant Sgr streams has been mis-identified, and 
that one or more of the `bifurcated' pieces is the real Sgr tidal tail, but we do not reach definite
conclusions.  
If two dwarf progenitors are assumed, fits to the planes of the dominant and `bifurcated' tidal tails favor an association of the Sgr dwarf spheroidal galaxy with the dominant Southern stream and the `bifurcated' stream in the North.
In the North Galactic Cap, the best fit Hernquist density profile for the smooth 
component of the stellar halo is oblate, with a flattening parameter $q = 0.53$, and a scale 
length of $r_0 = 6.73$.  The Southern data for both the tidal debris and the smooth component 
of the stellar halo do not match the model fits to the North, although the stellar halo is still overwhelmingly oblate.  
Finally, we verify that we can reproduce the parameter fits on the asynchronous Milkyway@home volunteer computing platform.

\end{abstract}


\section{Introduction} \label{intro}

The discovery that a significant fraction of the stars in the Milky Way stellar halo are 
organized into spatial substructures \citep{nyetal02} has transformed our model of that component
of the Milky Way.  Before this discovery, the stellar density of this component of our galaxy 
was routinely fit as a power law \citep{op75, psb91}
or deVaucouleurs profile \citep{d77,bs80}, but spheroid substructure \citep{fieldofstreams} observed 
with the Sloan Digital Sky Survey (SDSS; York et al. 2000)
showed that fitting these smooth density functions is difficult.  Previous to SDSS, the data used to 
constrain the spheroid density used typically hundreds of spheroid tracers in shallow surveys containing 
a much larger sample disk stars \citep{gr83,psb91}, surveys of rare objects such as RR Lyraes and globular clusters
\citep{op75,db78,v02}, or a small number of narrow, deep fields \citep{gr83,bs84,rymts96,rrc00,smrt02}.
The SDSS data has been used to constrain the shape of the spheroid \citep{jetal08}, but the authors removed resolved overdensities in order to fit a smooth model to a demonstrably lumpy population of stars.
See \citet{gwk89, m93, h08} for a more complete list of references.

Of the spheroid substructures, the most obvious and best studied is the Sagittarius dwarf tidal stream.  
The Sgr dwarf galaxy was thought to be in the process of tidal disruption from the time of its discovery \citep{igi94} due to the fact that it is elongated and only about 16 kpc from the Galactic center.  
The tidal stream was first discovered from some dozens of carbon stars \citet{ilitq01} on a great circle along the Sgr dwarf orbit.  
At nearly the same time, overdensities of blue horizontal branch and blue straggler stars were discovered \citep{ynetal00}, which turned out to be cross sections through the tidal stream \citep{iils01}.  
Although early calculations pointed to a large dark matter content of the Sagittarius dwarf \citep{iwgis97}, which was required for the dwarf galaxy to survive for the age of the Universe, one possible explanation
is that the dwarf galaxy was deflected into its current orbit within the last 3 Gyr \citep{z98}.
Later simulations by \citet{hw01} showed that the observations could be reproduced with or without a massive dark matter halo, even over 12.5 Gyr.
Since its initial discovery, many groups have found Sgr stars in the halo, but two of the most 
striking images are from the Two Micron All-Sky Survey \citep{mswo03} and the SDSS 
``Field of Streams" \citep{fieldofstreams}.

Many controversies have surrounded the Sagittarius dwarf galaxy.  Initially, the leading tidal tail of 
the Sgr dwarf appeared to arc through the North Galactic Cap and rain down on the position of the Sun 
\citep{mswo03,ljm05,mpjai07}, sparking a debate about the effects of a dark-matter dominated stream on 
direct detection of dark matter \citep{fgn05}.  
However, later work that traced the stream with more numerous color-selected F turnoff stars \citep{nyetal07} 
clearly showed that the leading tidal tail passes through the Galactic plane well outside the solar circle 
towards the anticenter.   
In \citet{fieldofstreams} it is shown that there is an apparent bifurcation in the Sgr leading tidal tail.  
Several papers have attempted to explain the bifurcation as multiple wraps of the same stream \citep{fellhauer} or dwarf galaxy rotation \citep{petal10}.  
Spectroscopy showed that the velocities and metallicities of the stars in each part of the leading tail bifurcation were similar \citep{ynetal09}.
However, \citet{ketal12} showed that the trailing tidal tail was also bifurcated and the photometric metallicity 
measurements were not the same, so they suggested that in both cases the ``bifurcation" could be due to a second stream whose progenitor is not the Sagittarius dwarf spheroidal galaxy.  
If the stars that were previously assumed to have been stripped from the Sgr dwarf galaxy actually originate from multiple galaxies instead, then it complicates claims of debris from multiple wraps of the stream around the galaxy \citep{ietal00,cbifv10}.
If true, this would not be the first time that a substructure was later discovered to be composed of more than one component.  For example \citet{g06,lijing12} show that the ``Monoceros Ring" near the galactic plane \citep{nyetal02,ynetal03} in the anticenter has multiple components.  
Furthermore, the Virgo Overdensity
\citep{vetal01,nyetal02,jetal08} appears to be a mix of several components \citep{vivas12}.  

Perhaps the best known conundrum is over what the Sagittarius dwarf tidal tails imply about the dark matter potential.
Stars that have been tidally stripped from dwarf galaxies and globular clusters are the only stars for which we have information about their locations in the past.  As such, there is promise that they could be used to constrain the
Milky Way's gravitational potential, which is dominated by dark matter.  
Semi-analytic n-body simulations, in which the Milky Way is modeled with a parameterized analytic potential and the dwarf galaxy is simulated with a set of equal mass particles, have been compared to the location of the Sgr tidal stream.  Early analyses argued for a roughly spherical dark matter distribution, since the tidal debris was largely confined to an orbital plane \citep{ilitq01,mswo03}.
However, the measured line-of-sight velocities of the leading tidal tail favor a prolate dark matter potential 
\citep{law04,h04,vzg05}; and from orbital alignment of the tails, \citet{jlm05} measure an oblate dark matter halo.  
\citet{ljm05} show that it is not possible to fit both angular positions and radial velocities of the leading Sgr dwarf tidal tail in smooth, axisymmetric models.
More recently, a triaxial dark matter halo has been shown to be consistent with all of the data in the Sgr dwarf tidal stream \citep{lmj09,lm10}, though this result has made some uncomfortable because the Milky Way disk in their best fit model is orbiting about the intermediate axis, and this particular triaxial halo configuration is disfavored by cold dark matter Galaxy formation models \citep{allgood06}.
Though the triaxial model is currently the best we have, it does not explain the ``bifurcation," and since dark matter halos are likely more complicated than have been modeled to date for the Milky Way, it is possibly not the only model that could fit the data.
From n-body cosmological simulations, for example, it seems the shape of the dark matter halo is in general complex, and is a function of radius and time \citep{bast05,tiss10}.

Tidal streams are useful for tracing the gravitational potential of the Milky Way even if the dark matter potential is complex, lumpy, or time-dependent.  
Several techniques for measuring the large-scale shape of the dark matter halo have been put forward \citep{eb11,vil11, jmsr99}.  
In addition to the Sgr dwarf tidal stream, the Milky Way halo potential has been fit using the GD-1 cold stellar stream \citep{wnzyb09,krh10}, the NGC 5466 globular cluster tidal stream \citep{lux2012}, and an attempt has been made to simultaneously fit multiple streams at the same time \citep{willettthesis}.  
Tidal streams are also being used to constrain the lumpiness of the Galactic halo \citep{jsh02,ilic02, sv08, yjh11, c09, c12}, though see \citet{kkbh10} for an argument that tidal tail substructure is not a good indicator of halo dark matter substructure.  
The study of the dark matter potential and substructure is clearly a young field, and competing factors such as halo shape, effect of substructure, disk shocking, and time the dwarf galaxy has been on its present orbit have not been completely disentangled.  
Confusion from multiple stellar streams in the same volume (and sometimes even at similar velocities) in the observational data has also hindered progress.

The Sagittarius stream and other density substructures in the Milky Way make it difficult to fit the shape of the stellar halo.  
In general, the shape is expected to be a combination of the stars that fell in at early times, and is spatially well mixed \citep{hss03}, plus a component including dwarf galaxies and their associated tidal debris that fell in at late times or is currently infalling, as we see in for example the Sgr dwarf tidal stream.  
One can see from \citet{nyetal02,fieldofstreams} that very little of the observed halo is free from the presence of very large density perturbations from tidal streams.  
To fit a shape, one either averages over the density substructures, thus making the model fit sensitive to the part of the stellar halo sampled, or uses the small region of the stellar halo that does not appear to contain a dominant substructure component to fit the shape of the stellar halo with large substructures removed.  

In this paper, we take the first step towards calculating a density model of the stellar halo for each stellar substructure, and then fitting a smooth density model to the part of the halo that is left, by measuring the density of the Sagittarius dwarf tidal stream in the North Galactic Cap.  Once the Sgr density is known, a smooth spheroid model can be more easily fit to the remaining stars.
We accomplish this using the statistical photometric parallax technique \citep{newberg13} in which the statistical properties of a stellar population are used to constrain the spatial density of those stars.

In order to separate the stream substructure from the smooth component of the stellar halo, a maximum likelihood algorithm was developed that found the best parameters, given the data, for a model consisting of the stream (or a number of streams) and a Hernquist model of the stellar halo \citep{cetal08,colethesis}.  
The data included the sky position and apparent magnitudes of color-selected F turnoff stars from the SDSS, chosen to be bluer than the turnoff of the thick disk.  
Because F turnoff stars have a wide range of absolute magnitudes, their distances cannot be determined individually; however, the absolute magnitude distribution can be used to statistically determine the underlying density distribution of stars.
Although previous descriptions of our fitting algorithm only used the data in one $2.5^\circ$-wide SDSS stripe at a time, we found it necessary paper to extend the algorithm so that information about the positions of the stream in neighboring stripes could be used in the optimization.

In this paper, we present the density distribution of Sagittarius dwarf tidal debris in the portion of the North Galactic Cap surveyed the SDSS, and in three $2.5^\circ$-wide stripes that were included in the sixth SDSS data release.  
A sample set of stars that has the density characteristics of the Sgr dwarf tidal stream is included as an electronic table.  
This table was generated by asking, for each star in the input dataset, what the probability is that it is in the Sgr stream, and selecting it with that probability.  
Note that although some of these stars are actually in the Sgr stream, the sample of Sgr stars does not in general indicate which stars are actually in the Sgr stream; they are only a set of actual stars that have the expected density characteristics of the stream.

We use the results of fitting the substructure densities to show that the widths of the dominant Sgr tidal stream are consistent with those of recent n-body simulations, but there is a discrepancy in the stellar densities.  
We show that the stellar halo is oblate, while the halo potential (including dark matter) is most consistent with a triaxial or axisymmetric prolate shape.
Additionally, we study the orbital planes of the separate Sgr stream components (the dominant streams and the bifurcated pieces) and explore the possibility that one or more of the dominant streams is not in fact tidal debris originating from the Sgr dwarf.  Finally, we verify our maximum likelihood parameter fits using the Milkyway@home computing platform, and discuss its future utility.

\section {Extraction of Data from SDSS}\label{datasetcreation}

Our data was queried from the SDSS photometric catalog, with data for the South Galactic cap 
taken from the DR 6 catalog \citep{DR6}, and data for the North Galactic cap taken from the DR 7 catalog \citep{DR7}.  
The SDSS data is organized into a series of great circle (GC) stripes that are 2.5$^{\circ}$ wide 
and varying length, typically 140$^{\circ}$ ($\nu$ and $\mu$ in SDSS GC coordinates, 
respectively).  This implies that each stripe produces a wedge-shaped volume of data, with 3-dimensional 
spherical axes consisting of apparent magnitude, and the two SDSS GC angles.  The model used in our maximum 
likelihood analysis takes advantage of the piecewise nature of the data by analyzing each stripe 
separately.  

All sources from a given stripe that were identified as point sources were selected to have the color 
of blue F turnoff stars.  It was shown in \citet{nyetal02} that the Sgr turnoff is much bluer than that of 
the thick disk stars and even slightly bluer than the stellar halo in general.  Selecting those stars 
consistent with the colors of the Sgr turnoff preferentially selects stars in the stream and helps 
to limit the amount of contamination from other stellar components of the Galaxy.

Those sources with colors $0.1 < (g - r)_0 < 0.3$ and $(u - g)_0 > 0.4$, that did not have EDGE or SATURATED flags
set, and with magnitude $g_0 > 16$, were included in the dataset for a given stripe number.  The subscript ``0," here 
denotes that reddening corrected magnitudes were selected.  All of the selected stars are have 
$\left | b \right | > 30^{\circ}$ and have distances $R > 2.3$ kpc.  All stars should then be well 
outside of the Galactic plane, and therefore past the vast majority of dust, so 
the \citet{sfd98} dust maps (included in the SDSS database) are used to
correct for reddening biases in the photometry.  The color cut in $(g - r)_0$ 
selects blue F turnoff stars in the halo, whose average absolute magnitude is
$M_{g_0} = 4.2$, as shown by \citet{nnscm11}.
The color cut in $(u - g)_0$ is performed in order to remove contamination by low redshift 
quasi-stellar objects (QSOs) \citep{nyetal02}.  Stars brighter than $g_0 = 16.0$ were removed to avoid 
saturated stars. At this magnitude and color most if not all contamination by disk stars is avoided, since magnitude 
$g_0 = 16$ corresponds to the approximate distance of $R = 2.3$ kpc (a miniumum of 1.15 kpc above the plane for stars
above $b=30^\circ$).  At bright magnitudes (close distances), the $(g-r)_0$ colors are sufficiently accurate that there should be few thick disk stars in the sample, even when we are close to the Galactic plane.  

The datasets are typically limited to $g_0 < 22.5$, since the efficiency function, described in section 3.1.3, begins to fall rapidly at this apparent magnitude. 
However, in some stripes the Sgr stream is far enough away that we need to use the fainter data ($g_0 \approx 23.0$), even though the efficiency function is low. 
The faint limit of the stripe is increased (see Table 1) in instances where the Sgr stream center is fainter than $g_0 = 22.0$.
Although the angular length of each stripe is typically $\Delta \mu \approx 140^{\circ} $, the actual
angular lengths used in our analysis were shorter in order to reduce the complexity of the model 
required to describe the data. For example, the Monoceros stream \citep{nyetal02,ynetal03} is seen towards the anti-center of the Galaxy and is apparent in the data along the edge of many stripes within the north 
Galactic cap. Therefore, the angular lengths of these stripes were chosen to specifically avoid
this structure. Due to the Galactic bulge, the Hernquist profile is often not a good fit 
to the data towards the Galactic center, so these sections were also removed.  A complete list of 
individual stripe bounds can be found in Table \ref{DataChar}.

We remove any section of a dataset (in $l$ and $b$) that contains a globular cluster or a region of poor data quality, 
and then the optimization over these datasets is performed using the algorithmic method for removing a 
section from the probed volume, as described in Section 3.  A list of cuts and their associated properties can 
be found in Table \ref{CutChar}.

The resulting dataset includes 1.7 million blue turnoff stars (selected to be bluer than the turnoff of the
Milky Way's thick disk) from eighteen SDSS stripes - fifteen of which are in the Northern Galactic Cap.
Table \ref{DataChar} describes the number of stars in each of the stripes.

\section{Algorithm for Fitting Tidal Debris}\label{algorithmforfittingtidaldebris}

The turnoff stars in each stripe of data were fit to a model of the spatial density of stars in the Milky Way stellar spheroid, including a smooth component and tidal streams, using statistical photometric parallax.
Since our data selection was designed to avoid other stellar components of
the Galaxy, we did not include density models for thin/thick disks, bulge, etc.
The basic algorithm for fitting spheroid substructure is described in detail in \citet{cetal08},
Section 2, and summarized here.  Extensions to that algorithm that allow us to fit multiple streams simultaneously are described in \citet{colethesis} and below.
Finally, we describe additional methods that used information from neighboring stripes to help
achieve the highest likelihood shape for the Sgr dwarf tidal debris.

A smooth component of the stellar halo, as we expect formed from early mergers that are well mixed, was fit with a Hernquist profile \citep{h90}:
\begin{eqnarray}\label{eq_spheroidModel}
\rho_{spheroid}(\vec{p}) \propto \frac{1}{r (r + r_0)^3}\mbox{, where}\\
r = \sqrt{X^2 + Y^2 + \frac{Z^2}{q^2}}. \nonumber
\end{eqnarray}
In this equation, $X, Y,$ and $Z$ are Galactocentric Cartesian coordinates, with $X$ in the direction from the Sun to
the Galactic center, $Y$ in the direction of the Solar motion, and $Z$ perpendicular to the Galactic plane pointing
towards the North Galactic Cap.  The stellar spheroid is thus defined by the two parameters, $q$ and $r_0$ (in kpc).  
The parameter $q$ measures the flattening of the stellar halo along the $Z$ axis.
The stellar spheroid could be oblate $(q < 1)$, spherically symmetric $(q = 1)$, or prolate $(q > 1)$.  The 
parameter $r_0$ is a core radius that determines at what distance from the Galactic center the Hernquist profile transitions
from $\rho \propto r^{-1}$ for small $r$ to $\rho \propto r^{-4}$ at large $r$.  The Hernquist profile was chosen because it is a reasonable guess for the density function of stellar halos \citep{bj05}, and
has a small number of parameters; however, this is not the only choice that could have been made.  
For example, we could have allowed triaxiality in our Hernquist fit, following recent work indicating a triaxial dark matter halo \citep{lmj09,lm10}.  However, our goal in this paper is to remove the Sagittarius dwarf tidal debris from SDSS F turnoff data, so that other components may be fit accurately in the future, and so we chose an axisymmetric halo density distribution with a minimal number of free parameters.  
\citet{cetal08} showed that the fit parameters of the tidal streams were not sensitive to the parameters of the smooth component.
A detailed fitting of the halo, including testing a triaxial spheroid, will be performed once major tidal debris has been accounted for, and will be the subject of a future paper.

Tidal streams are modeled piecewise, with a separate straight sections through each $2.5^\circ$-wide SDSS stripe.  
The density falloff with radius from the stream center is modeled as a Gaussian.  The density along the model stream is 
constant for each short section.  This is only a good approximation if the stripe width is small compared to the density 
variation along the stream and compared to its radius of curvature, and if the cross sectional density of the stream is
well approximated as a Gaussian.  \citet{cetal08} concludes that this piecewise approximation is adequate.

The parameters describing the individual pieces of the tidal stream are discussed in detail in \citet{cetal08}. To summarize, there is a unit vector that defines the axis of a stream cylinder, 
$\hat{a}$, specified by two angles ($\theta,\phi$), a point that is constrained to be along the centerline
of the data stripe ($\nu=0$) that is on the axis of the cylinder a distance R from the Sun and an angular position $\mu$ along the stripe, and the sigma of the Gaussian that describes the width of the stream.  
In addition to these five parameters, each stream needs a normalization parameter, $\epsilon$, that describes the fraction of the stars in the stripe that belong to the stream.  
In total, six parameters are needed to completely define a cylinder with Gaussian density fall off from the cylinder axis within a single $2.5^{\circ}$ wide SDSS stripe:  $\mu$, $R$, $\theta$, $\phi$, $\sigma$ and $\epsilon$.  Using these parameters, the stellar density of the stream at point $\vec{p}$ can be described as:
\begin{equation}\label{eq_streamModel}
\rho_{stream}(\vec{p}) \propto e^{-\frac{d^2}{2\sigma^2}}.
\end{equation} 
where $d$ is the distance to the center of the stream and is a function of apparent magnitude $m_g$.

The likelihood function used by the algorithm is more fully explained in \citet{cetal08}.  
The likelihood is the product of the probabilities of observing each of the stars in the dataset, given the model for stellar density and observational constraints.  
The probability of observing a star at a given position is called the probability density function (PDF).  
The PDF includes a weighted combination of the spheroid and stream densities, an absolute magnitude distribution for the observed F turnoff stars, and a detection efficiency function.

The weight, $\epsilon$ of a model component (stream to spheroid, in this case) is related to the fraction, $f$, of the stars present in that component relative to the entire dataset.  
The stream and spheroid fractions are defined as:
\begin{eqnarray}
f_{stream_i} = \frac{e^{\epsilon_i}}{1+\sum_{j=1}^k[e^{\epsilon_j}]} \\
f_{spheroid} = \frac{1}{1+\sum_{j=1}^k[e^{\epsilon_j}]} \nonumber
\end{eqnarray}
where $i$ and $j$ denote the $i^{th}$ and $j^{th}$ stream, respectively, of $k$ total streams.  
We use this weight definition, rather than the star fraction, as a fitting parameter because the individual $\epsilon$ values are free to vary from $\infty$ to $-\infty$ without artificial constraints.  
An $\epsilon_i=\infty$ (with other $\epsilon_j$ $<\infty$) implies that all of the stars belong to the $i^{th}$ stream, while an $\epsilon_i=-\infty$ implies that zero stars belong to the $i^{th}$ stream, 
and $\epsilon_i=0$ implies that the $i^{th}$ stream has the same number of stars as the smooth component of the stellar spheroid (since the numerator of $f_{spheroid}$ is fixed at 1).

The dataset includes the angular position in the sky $(l,b)$, and an apparent magnitude, $g_0$.  
Since the F turnoff stars have a range of absolute magnitudes, the apparent magnitude only implies an approximate distance.  
If we assume the absolute magnitude distribution of F turnoff stars in the sample is a
Gaussian function with mean absolute magnitude of $\bar{M}_{g_0} = 4.2$ and a dispersion of $\sigma_{g_0} = 0.6$ \citep{ny06},
then we can define a probability distribution for the distance to each star.  It is simpler to think of this the other way,
that we can convolve the expected stellar density from our model with the Gaussian, and then compare the data with the
expected apparent magnitude distribution given the model for the density and the model for the absolute magnitude distribution:
\begin{eqnarray}
\rho^{con}_{comp}(l,b,\mathcal{R}(g_0)) &=& \frac{1}{\mathcal{R}^3(g_0)}
\int_{-\infty}^\infty \mathcal{R}^3(g)\cdot
\rho_{comp}(l,b, \mathcal{R}(g_0)) \cdot
\mathcal{N}(g_0-g;u) dg , \nonumber
\end{eqnarray}
where $\mathcal{R}$ is the Sun-centered distance to the star implied by an intrinsic $M_{g_0} = 4.2$, $\rho_{component}$ is density function of the current model component (spheroid or stream), and $\mathcal{N}$ is the Gaussian density function given by:
\begin{equation}
\mathcal{N}(x;u) =  \frac{1}{u \sqrt{2\pi}} e^{\frac{-x^2}{2 u^2}},
\end{equation}
with $u=\sigma_{g_0}=0.6$.  Therefore $\rho^{con}_{component}$ is the stellar density of a particular component of our model, which takes into account the intrinsic magnitude spread of F turnoff stars.  

Since the SDSS is a magnitude-limited survey, the chance that a star will be detected decreases as its apparent magnitude 
increases to the limiting magnitude.  Therefore a detection efficiency function, $\mathcal{E}$, is defined which describes 
the percentage of stars detected at a given magnitude.  This function was illustrated in Figure 2 of \citet{nyetal02} and is
characterized by a sigmoid curve:
\begin{eqnarray}\label{eq_efficiencyFunction}
\mathcal{E}(g_0) = \frac{s_0}{e^{s_1 (g_0 - s_2)} + 1}\hbox{, where}\\
(s_0, s_1, s_2) = ( 0.9402, 1.6171, 23.5877 ), \nonumber
\end{eqnarray}
We incorporate this detection efficiency function into our model as a direct multiplier to the stellar density model, which 
causes the model density to gradually roll off at higher apparent magnitudes.

We can then write the final PDF, including the density distribution of stars in the spheroid and each stream, the absolute
magnitude distribution of F turnoff stars, and the fact that the dataset includes a smaller fraction of the stars near the
faint magnitude limit, as:
\begin{eqnarray}\label{eq_pdfComplete}
PDF &=& \sum_{i=1}^k 
\left[ f_{stream_i}
\frac{\mathcal{E}(\mathcal{R}(g_0)) \rho^{con}_{stream}(l,b,\mathcal{R}(g_0) | \vec{Q}_{stream_i})}{\mathcal{I}_{stream_i}^{com}}\right] \nonumber\\
& & + f_{spheroid}\frac{\mathcal{E}(\mathcal{R}(g_0)) \rho^{con}_{spheroid}(l,b,\mathcal{R}(g_0) | \vec{Q}_{spheroid})}{\mathcal{I}_{spheroid}^{com}},
\end{eqnarray}
where $i$ and $j$ denote the $i^{th}$ and $j^{th}$ stream of $k$ total streams; $\mathcal{I}$ denotes the integral of the density over the entire volume for the respective component, corrected for volume cuts; 
$l$, $b$, $\mathcal{R}(g_0)$ are the input data; and $\vec{Q}$ is the set of parameters for the respective component.

Our complete PDF, then, contains $2 + 6k$ parameters:  2 for the spheroid ($q$, $r_0$), and 6 for each of $k$ streams ($\epsilon_i$, $\mu_i$, $R_i$, $\theta_i$, $\phi_i$, $\sigma_i$).
 
To find the model parameters that best fit the data (that is, the parameters that have the maximum likelihood), we use a combination of Conjugate Gradient Descent (CGD) and Line Search techniques.  
A technical description of these methods and validation tests can be found in Appendices A1 and A2 of \citet{cetal08}. 
Since the CGD algorithm is iterative and local, it is unaware of the global nature of the likelihood surface and cannot differentiate between a local maximum likelihood and the global maximum likelihood.  
Therefore, for a single given starting position of our search, we cannot be certain that the returned solution is the true global best fit parameter set.  
To compensate for these shortcomings, we chose starting parameters that were close to what appeared to be the best solution (``by eye"), then produced ten random parameter sets near those values\footnote{We chose ten runs per iteration because this was a convenient number of simultaneous runs to use with our available computing time in the queue system; additional starting points were generated based on the results of these ten (see next paragraph).  See Section 10 for a discussion of our computing resources, and verification of our results with a global search method.}.
Each of these parameter sets were then used as the starting values for a single execution of the algorithm.
Additional random input parameter sets were then generated about the best final fit parameters of the initial runs, and then the algorithm was run again.  
The final fit parameters of the run that produced the greatest likelihood were taken to be the true best fit.  
Through this method, we were able to avoid local maxima and find the global best fit of our model.

For the adjoining stripes 9-23 in the North Galactic Cap, we can impose the additional constraint of requiring the Sgr stream to be continuous at the stripe boundaries.  
When the random starts produced results that were discontinuous at the boundaries, new start positions were generated based on the best fit parameters of the adjacent stripes.  
The stream direction and center was placed along the line connecting the two adjacent stripes' stream centers, and the width was the average of the two adjacent stripes' stream widths;  
and all other parameters were started at the current best fit values for the current stripe number.  
These new parameters sets were then used as initial parameters for the CGD algorithm, and were allowed to converge normally; that is, these initial parameters were still allowed to vary during the fitting process.  
Ten runs per stripe were then run at the same time, with slightly perturbed widths ($\sigma$), 
weights ($\epsilon$) and central stream positions ($R$ and $\mu$).  
This method could not be applied to the end stripes (9 and 23) since they had only one adjacent stripe to compare against;  these stripes were left at their best fit values from the initial search.  
When this process produced a better likelihood for an individual stream then the parameters for that stripe were updated.  Using this method, the stream centers shifted slightly, so this was an iterative process, calculating new stream directions from adjacent stripe centers until the likelihood did not improve for any stripe.
This iterative process was necessary because the likelihood surface in the vicinity of the best parameters was fairly flat, so the conjugate gradient search often terminated due to slow progress near the best or was caught in a local maximum.

The results presented in Section 5 are those from the runs that produced the best likelihoods out of any of the above methods.

An estimate for the variance of each parameter can be found from the square matrix of second-order partial derivatives of the likelihood function, evaluated at the convergence value of the maximum likelihood search.  
This matrix is called a Hessian Matrix, $\mathbf{H}$.  The variance matrix $\mathbf{V}$ is the inverse of the Hessian Matrix, normalized by the number of stars in the data set:
\begin{equation}
\mathbf{V} = \frac{1}{N} \mathbf{H}^{-1}.
\end{equation}
The statistical error in the measurement of each parameter is estimated from the square root of the diagonal elements of the matrix.

The step sizes for calculating the Hessian matrix were chosen to be of the size of the expected errors in the fits.
After calculating the errors with these step sizes, some of the errors would be nonsensically large or small.  
In these cases, the step size for that parameter was increased by 50\% and the Hessian matrix recalculated.  
This process was continued until the Hessian error was within 50\% of the value of step size used.  
This method was necessary, as the likelihood surface (eight- to twenty-dimensions) is very flat near the maximum.  
Random fluctuation can therefore be very significant when determining the best Gaussian fit to the peak, if the peak is sampled in the flat region.
By slowly increasing the step size, we eventually step out of the noisy area without stepping away from where the curvature is relevant to the best-fit location.

\section{Dividing the Dataset Between Multiple Tidal Debris Streams and the Spheroid}\label{separatingdebris}

Given a dataset and best-fit model parameters, it is possible to separate the data into 
subsets that have the density profile of each component that was fit during optimization (in this case a spheroid component with a Hernquist profile plus one or more stream components).  
We do this by populating catalogs of independent structures in such a manner that the density distribution of that structure is accurately represented; however, it is not possible to populate a catalog for a given structure with only stars that are physically from that structure.  

For each star, the probability, $T_i$, that it is drawn from the $i^{th}$ stream population is calculated, given the parameters, by:
\begin{equation}\label{eq_separation}
T_i(l,b,\mathcal{R}(g_0) \mid\vec{Q}) = \frac{S_i(l,b,\mathcal{R}(g_0)\mid\vec{Q})}
{S(l,b,\mathcal{R}(g_0)\mid\vec{Q}) + B(l,b,\mathcal{R}(g_0)\mid\vec{Q})},
\end{equation}
where 
\begin{equation}\label{eq_separationSpheroidPart}
B(l,b,\mathcal{R}(g_0)\mid\vec{Q}) = f_{spheroid}\frac{\rho^{con}_{spheroid}(l,b,\mathcal{R}(g_0) | \vec{Q}_{spheroid})}{\mathcal{I}_{spheroid}^{com}}
\end{equation}
and
\begin{equation}\label{eq_separationStreamPart}
S(l,b,\mathcal{R}(g_0)\mid\vec{Q}) =  \sum_{i=1}^k \left[f_{stream_i} \frac{\rho^{con}_{stream}(l,b,\mathcal{R}(g_0) | \vec{Q}_{stream_i})}{\mathcal{I}_{stream_i}^{com}}\right].
\end{equation}
$S_i(l,b,\mathcal{R}(g_0)\mid\vec{Q})$ is defined as the $i^{th}$ component of the sum in equation~\ref{eq_separationStreamPart}.

After probability, $T_i$ is calculated it is tested against a random number, $\lambda$, generated uniformly on the interval 0 and 1.  
The star is placed into the $i^{th}$ stream catalog with probability equal to $T_i$ and the spheroid catalog with probability $1-\sum_1^k{T_i}$.  
Thus, distinct population catalogs are created for each structure over which we optimized:  one for the smooth spheroid, and one for each tidal stream.

This separation technique allows for the effectiveness of the model fits to be analyzed visually, and the separated catalogs provide density information that will prove useful for fitting tidal disruption models.  
The structure catalogs should not be used for selecting spectroscopic follow-up targets to further study the characteristics of stars drawn from a specific population (composition and velocity data, for example);  these follow-up targets would be better selected through the explicit use of the probability, $T_i$, that a star is drawn from that population, rather than on the catalog which is a random assignment based upon this probability.

\section{Fitting the Sagittarius Tidal Debris}\label{sgr-section}

The fitting technique described above is applied to the characterization of the Sgr tidal tails in northern stripes 9-23, and southern stripes 79, 82, and 86.  We assume the Sun is 8.5 kpc from the Galactic center throughout our analysis.  

While the Sgr tidal stream is the dominant over-density in the Galactic halo, it is not the only significant structure in the SDSS data, as is commonly illustrated by the ``Field of Streams" image \citep{fieldofstreams}.  
In addition to the main Sgr stream, a secondary, fainter (less-dense), roughly parallel structure can be seen; this is has been dubbed the ``bifurcated" piece of Sgr \citep{fieldofstreams}.  
This fainter stream is a source of confusion for our algorithm, especially when it coincides with the angular coordinates of the Sgr stream.  Also located in SDSS, and close to Sgr, is the Virgo stellar stream \citep{vetal01}.  
Virgo is at lower Galactic latitude than Sgr, and represents a significant over-density in stripes 9-12.  
At low Galactic latitudes, in the opposite direction of the Galactic center (the anti-center) lies the Monoceros over-density, as well as other, fainter structures \citep{lijing12}.  These low-latitude structures are easily avoided by ignoring data at low $|b|$.  
In the South Galactic cap, the Hercules-Aquila cloud \citep{herc-aquil} represents the only (known) significant structure in our data.  This structure is easily avoided, as it is well-separated from the Sgr detections in the Southern stripes.  
Finally, there may exist other unaccounted-for substructure, data artifacts, or non-uniformities in the data. 

Other than Sgr and the over-densities that can be easily avoided (by selecting only sections of the data containing the Sgr stream), we are left with the faint (or bifurcated) Sgr stream, the Virgo stream, and any miscellaneous non-uniformities that may exist in the data.  
In order to prevent these objects from contaminating our fits to the Sgr stream, we chose to fit 3 streams to stripes located in the North Galactic cap:  the first stream fits Sgr, the second stream fits either the bifurcated piece or Virgo, and the third stream could pick up any remaining non-Sgr over-density. 
 In the event that too many streams were being fit to the data, testing showed that the algorithm would reduce the $\epsilon$ of the extraneous streams to a large negative number, placing zero stars in that stream.  Therefore, allowing the algorithm to fit one or two additional streams would not affect the accuracy of the method, but does increase the time the algorithm takes to converge.  
The non-Sgr components were not subject to stripe-to-stripe continuity restrictions to allow them to adapt to the potentially different conditions in each stripe.  Since our main concern was fitting the Sgr stream, we do not analyze these additional components and discuss them only briefly.

The data in the South Galactic Cap appeared to be less cluttered than in the North, and so we fit only one stream to these stripes.  Noticeable structures in the South (such as Hercules-Aquila) lie far enough from Sgr that additional fitting streams were not required in order to cleanly detect Sgr.  However, the fainter, bifurcated piece in the South was not accounted for in our fits.  It is possible that this bifurcated piece was included in the main Sgr stream by the algorithm;  if so, it may have skewed the fit to have a larger width or gave it a slightly different direction.  It is also possible that the bifurcated piece in the South is too faint for our algorithm to detect.  In any case, the separation plots (described in the next section) for the Southern stripes appeared to cleanly remove all over-density, so we believe that our fits are reasonable.  We will re-analyze the Southern stripes when we apply our methods to SDSS DR8 in the near future.

Using the algorithm outlined in previous sections, we found the best-fit Sgr stream parameters, which are tabulated in Table 3.  The $\mu$,$R$ parameters define the spatial location of the Sgr stream center, and the $\theta$,$\phi$ parameters define the stream direction.  The $\epsilon$ parameter describes the weight of the stream relative to the background, and $\sigma$ gives the standard deviation of the stream width.  `\# Stars' is the number of stars that are consistent with the Sgr stream in a given stripe.  These are depicted in Figures 1 and 2, which show the positional and directional results of optimizations on all of the SDSS data analyzed here.  The fits describe a smooth orbit from stripe 23 to the Sgr core and around the trailing tidal tail.  Figure 4 of \citet{nyetal07} performed a similar study using F turnoff stars, by fitting the the turnoff of the Sgr stream.  The results presented in Table 1 of \citet{nyetal07} are, in general, very similar to this paper in Galactic longitude and latitude determinations for each of the stripes.  However, the distances for stripes 15-22 are on average 2 kpc larger than those produced here by the maximum likelihood method.

A possible explanation for this discrepancy could be that the fits performed in \citet{nyetal07} were simultaneously fitting the main Sgr stream as well as its bifurcated section.  This would have the effect of moving the center of the detected structure further away, caused by the stream being much wider due to the combination of two distinct pieces of the stream.  
This effect could also explain the differing Galactic longitude determinations as well, for fitting the Sgr stream and its bifurcated section simultaneously at a larger distance would affect the determination of the location of the stream center.  
The stream appears to be well fit in both cases, however; this exemplifies the difficulty and need for determining a consistent definition of the ``center" of a detection.  

The stream centers have been converted into common coordinate systems in Table 4:  Galactic Cartesian $XYZ$, centered on the Sun with the Galactic center in the positive $X$ direction;  Sun-centered Galactic longitude and latitude, $l,b$; the Sun-centered Sagittarius $\Lambda_{\odot}, B_{\odot}$, coordinate system first described in \citet{mswo03}, using the C++ code made publicly available by David Law; and a new Sun-centered Sagittarius $\Lambda_{\rm new}, B_{\rm new}$ that reflects the results of this paper, defined in Appendix A.2.  

\begin{figure}[t!]
\label{plane_plots}
\figurenum{1}
\includegraphics[width=1.0\textwidth]{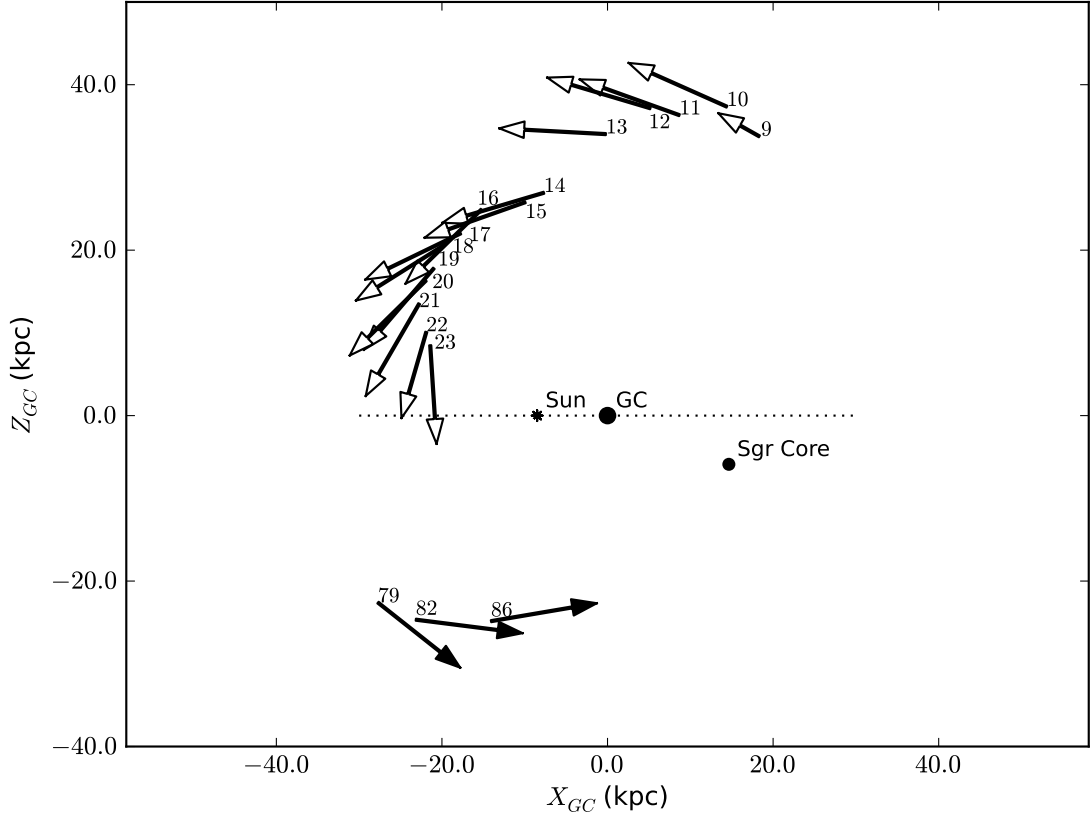}
\caption{\it Sagittarius stream fit directional vectors plotted in Galactic $X,Z$.  The base of each arrow is labeled with its respective SDSS stripe number.  White arrows represent stream detections located above the disk of the galaxy, while black arrows represent stream detections located below the disk.  The base of each arrow represents the location of the stream detection, while the head of the arrow illustrates the direction of the axis ($\hat{a}$) of the cylindrical shape that is fit to the stream in that stripe.  The length of the arrows is set by the projection of the unit vector $\hat{a}$ in the $X_{GC}, Z_{GC}$ plane, each scaled by the same arbitrary multiple to make the relative lengths more apparent.  The dotted line shows the plane of the Milky Way.}
\end{figure}
\clearpage

\begin{figure}[h!]
\label{plane_plots_xy}
\figurenum{2}
\includegraphics[width=1.0\textwidth]{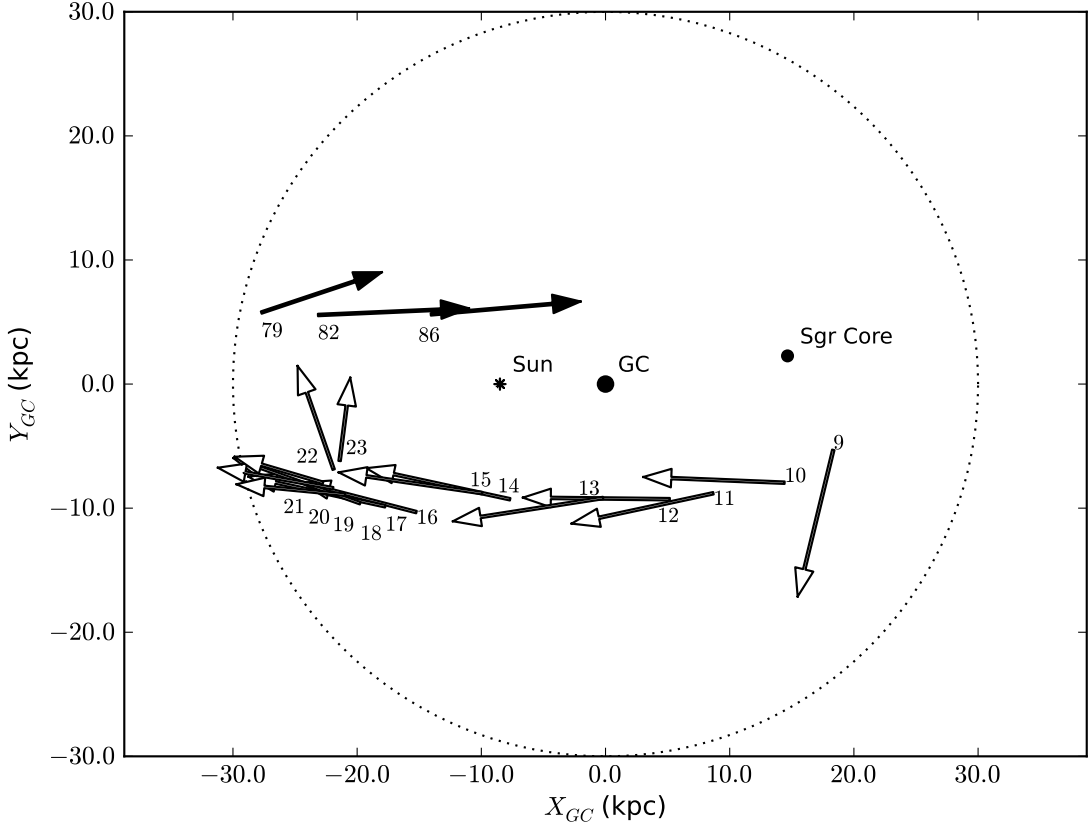}
\caption{\it Sagittarius stream fit directional vectors plotted in Galactic $X,Y$.  The base of each arrow represents the location of the stream detection, while the head of the arrow illustrates the direction of the stream.  The length of the arrows is the same arbitrary multiple of the unit vector $\hat{a}$ projections used in Figure 1.  The dotted line represents a 30 kpc ring centered on the Galactic center, roughly the edge of the Galactic disk.  
Note that the direction vectors are less accurate at the edges of the Northern data (SDSS stripes 9-23), where no adjacent stripes are available to guide the fit.}
\end{figure}

\section{Separating Sgr Tidal Debris from other Halo Stars}

Using the separation method described in Section~\ref{separatingdebris}, we were able to extract sets of stars that are consistent with the density profile of our maximum likelihood fits to the Sgr tidal stream.  
For each stripe, we would start with the original dataset of F Turnoff stars in $\mu,\nu,g_0$, as illustrated, using stripe 18, in Figure 3.  Using the separation method, we could separate the original data into subsets - one subset for each fit component: one subset for the background, and one more subset for each stream.  

\begin{figure}[!t]
\label{wedge_plot}
\figurenum{3}
\begin{center}
\includegraphics[width=0.6\textwidth]{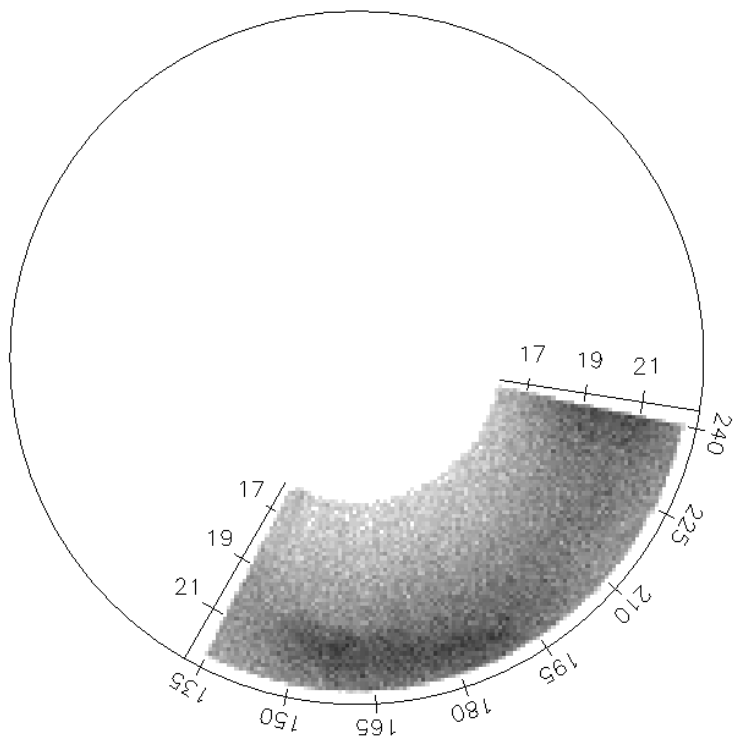}
\end{center}
\caption{\it Face-on wedge plot of SDSS Stripe 18 stellar density, in $\mu$ (angle along stripe as marked on the outside circle) and SDSS $g_0$ magnitudes (radial from center).  The radial position is proportional to $\log_{10}$ of the distance to the Sun.  Tidal debris can be seen as a dark density `cloud' in the region bounded approximately by $20 < g_0 < 22$, $150^{\circ} < \mu < 190^{\circ}$.  The density increases towards $\mu = 240^{\circ}$, since this direction passes closer to the Galactic center.  This stellar density wedge plot is typical of the SDSS data stripes analyzed in this paper.}
\end{figure}

\begin{figure}[!p]
\label{sep_plot}
\figurenum{4}
\begin{center}
\includegraphics[width=0.9\textwidth]{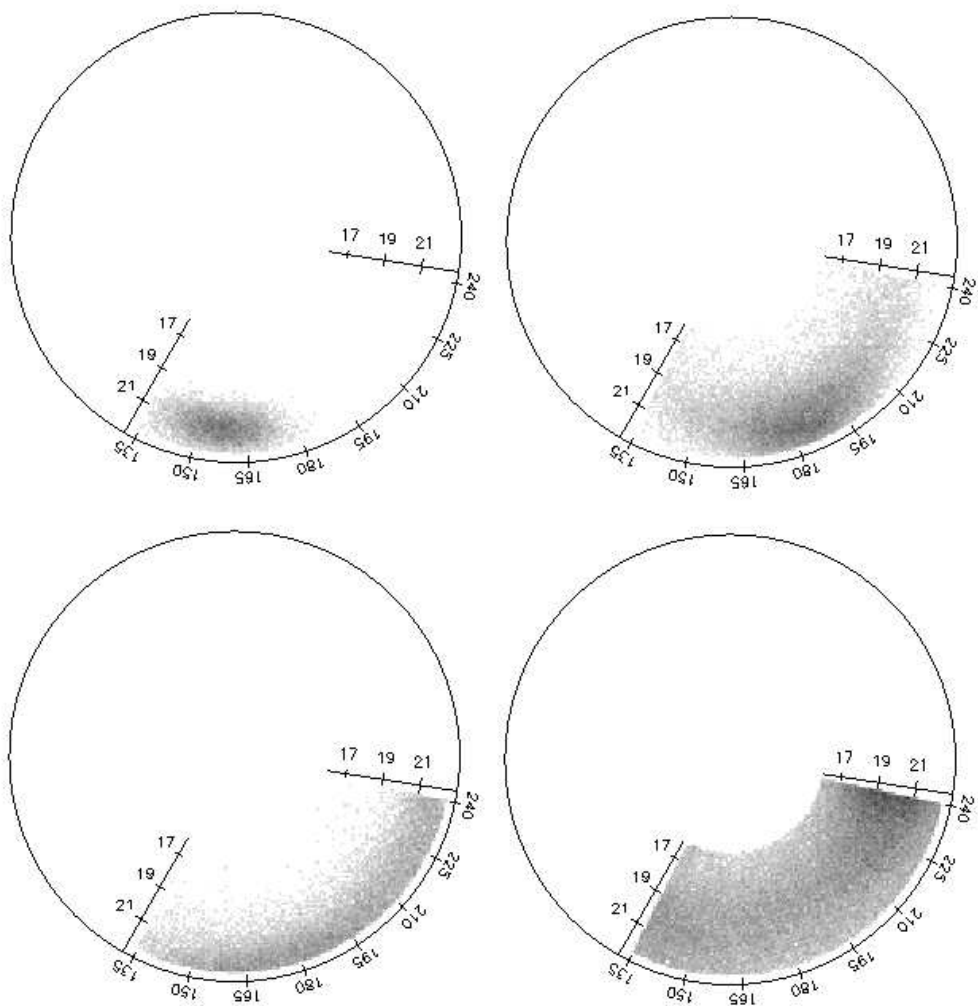}
\end{center}
\caption{\it Wedge plots of the probabilistically separated SDSS Stripe 18, with $g_0$ plotted radially, and the angle along the stripe, $\mu$, plotted in angle around the center in each panel.
Top-left:  Separated Sgr Stream, as fit within SDSS Stripe 18 data.  Top-right:  Secondary stream fit, most likely the bifurcated piece.  Bottom-left:  ``Garbage collection" stream.  Bottom-right: remaining smooth component, after fit streams have been removed.  Substructure appears to have been completely removed, leaving only the smooth component behind (note that the apparent over density near $\mu=240^{\circ}$ is the contribution from the Galactic bulge).  The segment of the Sagittarius tidal stream that passes through Stripe 18 also appears to have been cleanly separated.  The other two streams were not subjected to the stripe-to-stripe continuity requirements that was enforced for Sgr, and so were free to fit any density in the data not described by Sgr or the background Herquist profile.}
\end{figure}

Because of the additional substructure present in the Northern data, a single-stream fit would inevitably produce a questionable set of parameters for the Sgr stream, as it would attempt to fit the other substructure present in each stripe simultaneously with Sgr.  
We found that by fitting 3 streams to each North stripe, one stream would fit Sgr well, one stream would attempt to fit additional major substructure present, and the third stream would perform ``garbage collection" - picking up any density not described by the Hernquist profile, the Sgr stream, or other major substructure.  
The ``garbage collection" stream would always move out to outer $r$ limit of the data, and grow to be very wide.
We believe that this indicates that the Hernquist model does not sufficiently describe the smooth component of the stellar halo, and so the ``garbage collection" stream was needed to compensate for the Hernquist model's shortcomings.

We illustrate the separation process in Figure 4 by simultaneously plotting the $\mu,g_0$ subsets of 
stripe 18.  Here, $\mu$ is the angle along the stream in SDSS Great Circle coordinates.  The subset 
corresponding to the Sgr stream can be seen as the top-left plot, while the smooth Hernquist 
background can be seen as the subset plotted in the bottom-right.  Note that there is still an 
apparent over density near the edge of the data; this is the contribution to the background Galactic 
bulge.   
The very large, distant``garbage collection" stream is shown as the bottom-left dataset.  
The final subset, at top-right, is the rough fit to substructure secondary to Sgr;  
in stripe 18, this is most likely to be the so-called bifurcated piece.  The smooth appearance of the
remaining background is confirmation that the three streams have completely fit any non-smooth structure.

We present the entirety of the North Galactic cap data in an $l,b$ polar plot in Figure 5.  The North Galactic pole is located at the center of the plot, with Galactic latitude ($b$) decreasing radially.  This plot is comparable to the ``Field of Streams" plot, with the Sgr stream, its bifurcated piece, and the Virgo stream all clearly visible.

After extracting the stars consistent with the Sgr tidal stream from each SDSS data stripe, we were able to combine them to produce the complete Northern profile of the Sgr tidal stream, as seen in SDSS data, in Figure 6.
The stream is moving away from the Galactic center ($l=0^{\circ}$), and therefore the Sgr core, and becomes thinner and less dense the further it is traced from the core. The width appears constant along the stream; this is an illusion caused by the stream being more distant closer to the Galactic center, and becoming closer as it travels towards the edge of the galaxy.
A catalog of stars fitting the density profile of Sgr can be found in Table 5.

The stars remaining after the removal of the Sgr leading tidal tail can be seen in Figure 7.  Note that Sgr is cleanly removed (compare with Figure 5), leaving a smooth background in its place.  The bifurcated piece can still be found close to the center of the plot, and Virgo can be seen near the bottom of the data, near $l=300^{\circ}$.  The smooth density of stars increases toward the Galactic center, below $b=45^{\circ}$. 
We do not analyze these components further in this paper, as they were not subject to the rigorous fitting procedure used to derive the model fits to the Sgr debris, and will be briefly discussed in a later paper.

\begin{figure}[!t]
\label{polar_all}
\figurenum{5}
\includegraphics[width=1.2\textwidth]{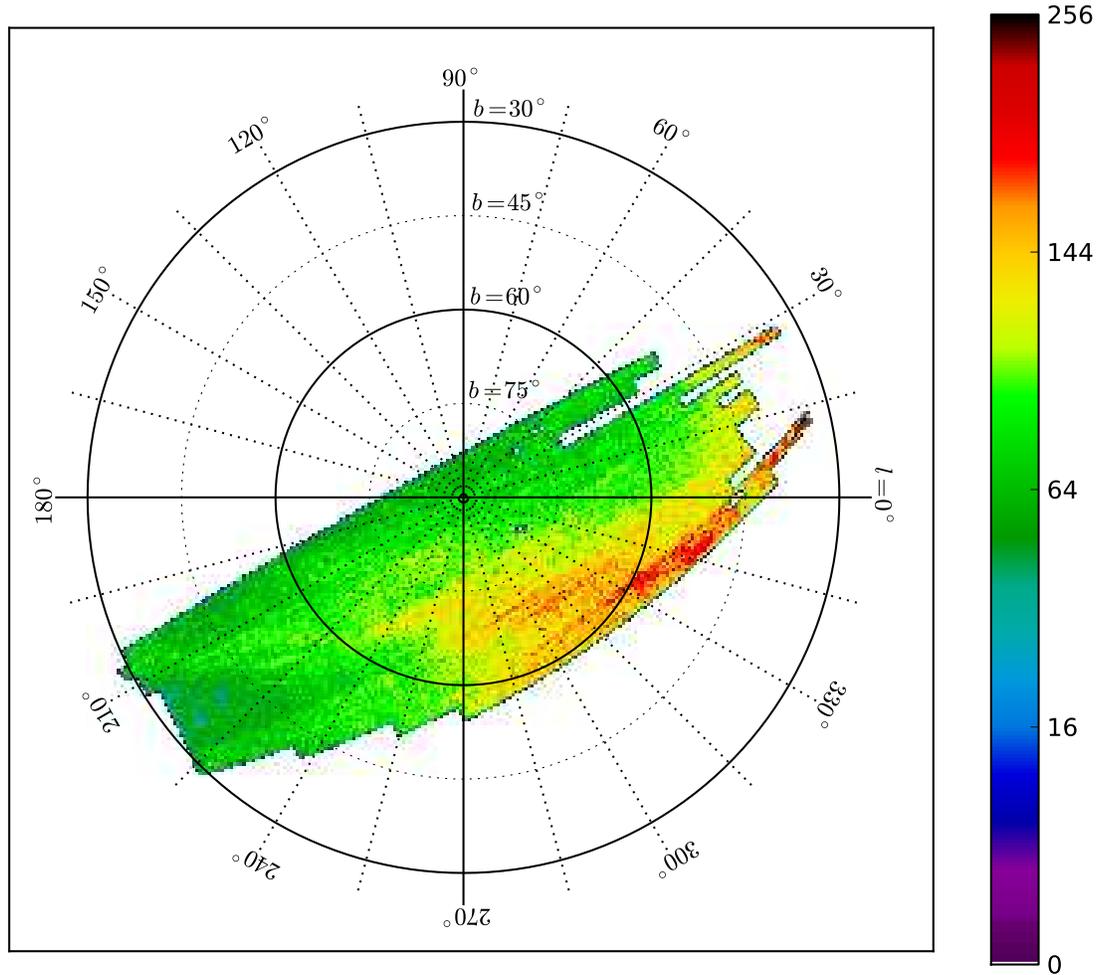}
\caption{\it Polar plot of F turnoff stars in the Northern Galactic Cap in Galactic $l, b$ coordinates.  
Each colored pixel is 0.5 degree by 0.5 degrees in size, with color indicating the total number of F 
turnoff stars in that area of the sky.
The Sagittarius stream is the overdensity near the center of the plot, spanning from $l,b \approx (210^{\circ}, 40^{\circ})$ to $l,b \approx (345^{\circ}, 55^{\circ})$.  The bifurcated piece can be seen as a faint secondary stream just above Sagittarius (closer to $b=90^{\circ}$).  The Virgo over-density is lower in the plot, and is centered (in the data) near $(l,b) \approx (315^{\circ}, 60^{\circ}$).}
\end{figure}

\begin{figure}[!t]
\label{sgr_plot}
\figurenum{6}
\includegraphics[width=1.2\textwidth]{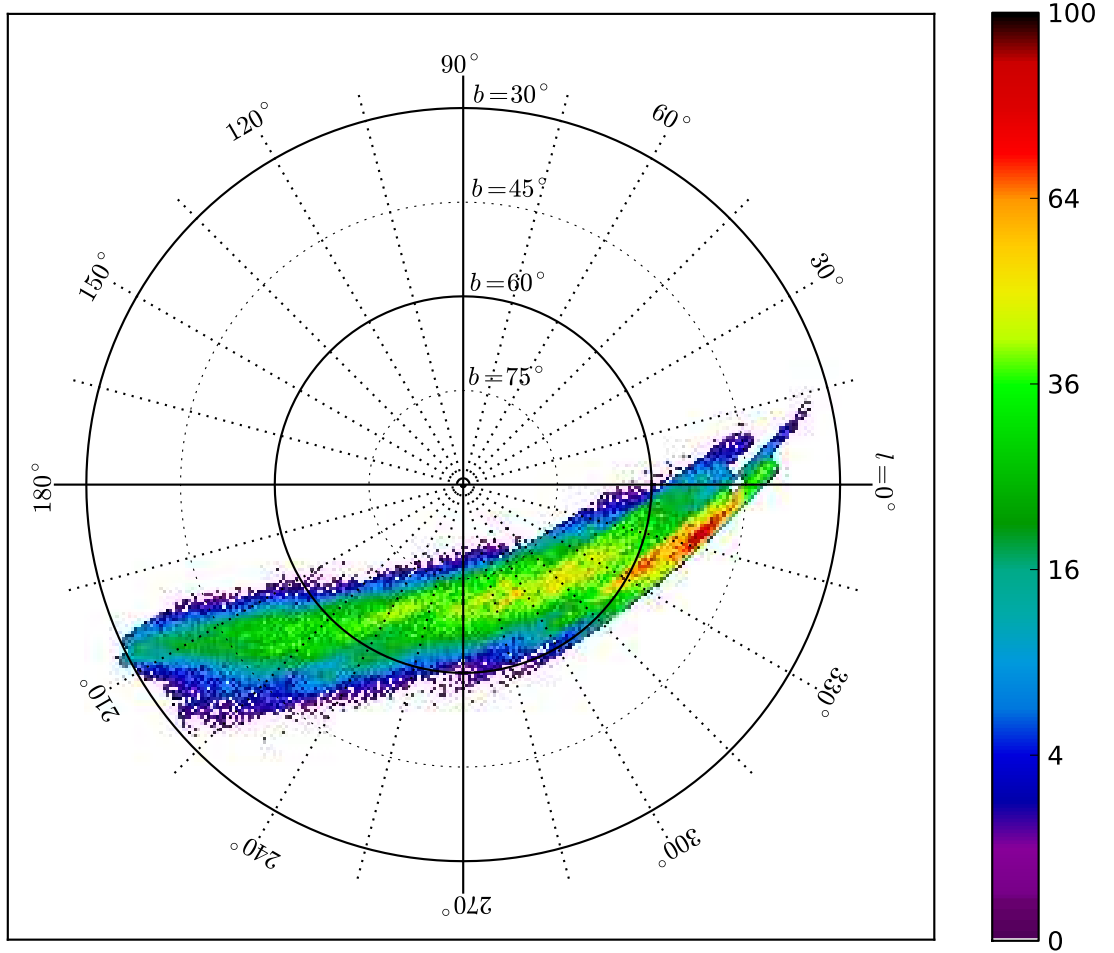}
\caption{\it Polar plot, similar to Figure 5, but with only the separated Sagittarius Stream shown.  The separated stream has a smooth, continuous distribution, despite having been fit to 15 separate SDSS stripes.  }
\end{figure}

\begin{figure}[!t]
\label{bkg_plot}
\figurenum{7}
\includegraphics[width=1.2\textwidth]{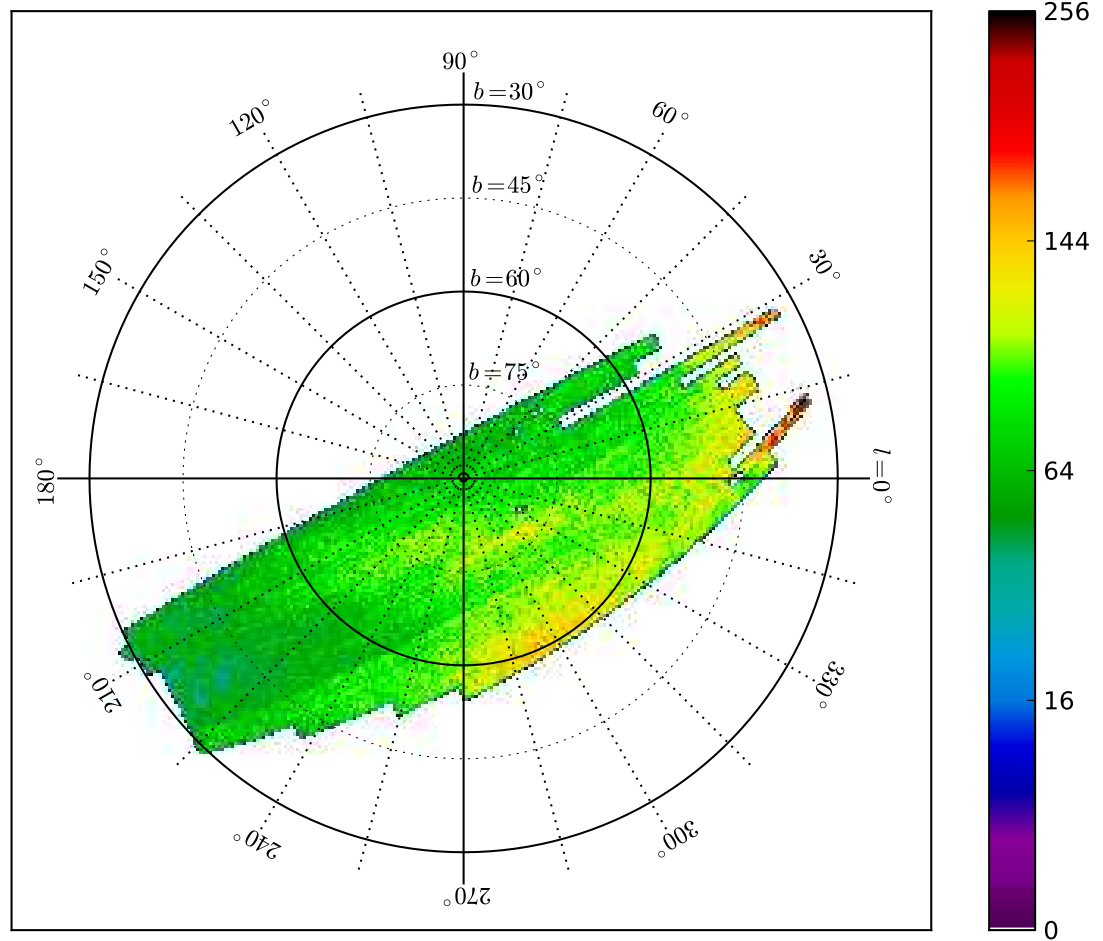}
\caption{\it Polar plot, similar to Figure 1, but with the Sagittarius Stream removed.  The bifurcated piece is clearly visible near the center of the plot, and the Virgo over-density is visible near the bottom of the data.  The Sgr Stream has cleanly been removed, leaving only Virgo, the Sgr bifurcated piece, and a smooth background.}
\end{figure}

\clearpage
\section{Plane Fits to the Sgr Tidal Tails}\label{sgr-planes}	

We find from the data in Figures 1 and 2 and Table 4 that the detections of the leading tidal tail lie in a plane, as does the trailing tail, but they do not lie in the same plane \citep{jlm05,ljm05,ketal12}.  The stream centers are also not consistent with the original orbital plane of \cite{mswo03}, which has equation\footnote{Note that we present this equation in right-handed $X,Y,Z$ coordinates, giving it a slightly different appearance from the left-handed \cite{mswo03} presentation.}
\begin{equation}\label{orbitalPlane}
-0.064X + 0.970Y + 0.233Z + 0.232 = 0.
\end{equation}  

We seek to define new orbital planes, one for the leading tail and one for the trailing tail.  Using all fifteen detections of the leading tail, a gradient descent method was applied to find the best fit plane to the points along the leading tail (See Appendix~\ref{app-A} for details, including error analysis.  A new set of Sun-centered in-plane $\Lambda, B$ coordinates is also defined, but not used in this paper outside of Table 4).  The resulting plane has the equation
\begin{equation}\label{newOrbitalPlane}
-0.199X + 0.935Y + 0.293Z - 0.988 = 0 
\end{equation}
with a correlation value of 0.019.  

The gradient descent method was again applied, this time to the three trailing tail detections, to find the best-fit plane to Sgr South, resulting in the plane equation
\begin{equation}\label{eq_orbitalPlaneSouth}
0.032X + 0.987Y + 0.158Z - 1.073 = 0 
\end{equation}
with correlation value 0.011.  

We choose a convention in which the $Z$-component of the plane normals (which are given by the coefficients 
of $X,Y$ and $Z$ in the above equations) is positive.  The normals to the two planes of equations \ref{newOrbitalPlane} 
and \ref{eq_orbitalPlaneSouth} are approximately $15.6^{\circ}$ ($\pm 0.1^{\circ}$) apart.  The leading and trailing tail appear to lie along differing orbital planes with the Sgr core lying -3.5 ($\pm 0.7$) kpc away from the leading orbital plane (that is, opposite the plane normal) and 0.7 ($\pm 1.0$) kpc from the trailing orbital plane.  The Galactic center lies -1.0 ($\pm 0.3$) kpc from the leading plane, and -1.1 ($\pm 0.5$) kpc from the trailing plane.  
The Sgr core is in better agreement with the trailing tail orbital plane than the leading tail orbital 
plane, which is expected, as the Sgr core and trailing tail are both located below the disk, 
and the trailing tail debris is closer to the dwarf and therefore has not yet evolved in the Galactic potential for as long as the leading tail.
Both orbital planes pass within the same distance of the Galactic Center ($\sim$1 kpc), but do not pass through it.

The fit planes could have been forced to pass through the Sgr dwarf core or the Galactic center, but our aim was to characterize the Sgr tidal debris; forcing the planes to intersect either of these points requires presumptions about the halo potential and stream associations.
By ignoring the Sgr core during the plane fitting process, we look only at the planes implied by our 
analysis, and discuss association with the Sgr core below.
Also, a recent study by \citet{ketal12} indicates that one or more of the tails and associated
bifurcations may not belong to Sgr; if this is the case, then including the core in the plane fits would not be appropriate.  
If we required the fits to pass through the Galactic center, we would be assuming that the Galactic halo is spherical.  
Given that the current best fit dark matter halo is triaxial \citep{lm10}, a symmetric orbit seems unlikely.  
In the past, the deviations of the leading and trailing tidal streams from a common plane were used as evidence for a non-spherical Galactic potential.  

Although we have plane fits for Sgr in the North and South Galactic Caps, it is instructive to compare the angles between planes with fits to the bifurcated Sgr pieces.  We use results from \citet{ketal12} to get an alternate fit to Sgr South and a fit to Sgr South bifurcation, using sky positions from their Table 1 and distances (corrected with recent Errata) from Table 2 of that paper.  We extrapolated distances using their magnitude gradient (see Figure 5 of that paper) where necessary.  We also include the plane fit from \citet{mswo03} in our analysis.  The best-fit plane parameters are given in Table 6, and the angles between respective plane pairs are presented in Table 7.  

We find from Table 6 that, of all the North-South plane pairs, the Sgr North and Sgr South planes from this paper are the farthest apart (15.6$\pm$0.1), while the two bifurcated pieces (Newberg 2007, Koposov 2012) are the most similar (3.9$\pm$2.2).   
Note that the main Sgr South orbital plane for {Koposov 2012} is nearly identical to that of \citet{mswo03}.  

There are three possibilities here:  (1) Sgr and the bifurcated pieces are from one dwarf galaxy, being described by multiple wraps, internal dynamics of the Sgr dwarf (rotation, etc.), or significant lumps in the halo; (2) The remains of two dwarfs are present, with some combination of leading-trailing streams describing both; or (3) tidal debris from more than two dwarf galaxies are present.

If (1) is true, and the two bifurcated pieces are from the same orbital wrap, then the low angle between their respective orbital planes is indicative of a symmetric, well-behaved halo potential.  
However, the large angle between the main Sgr stream orbital planes then implies a strongly asymmetric halo potential; both of these cannot be true.  Dynamics internal to the dwarf are still a possibility, but would have to describe a bifurcated piece that appears on the same side of the main stream both above and below the Galactic plane.  Several authors \citep{fellhauer,lmj09,petal10,lm10} have proposed mechanisms that would allow the Sgr streams to be a single continuous object, but no single theory has been able to completely explain all of the data.

If there are two dwarf progenitors (case 2), then we do not want to minimize the difference between two orbital plane angles, but rather to find the two North-South orbital plane pairs that have the most consistent (similar) angle differences and distances from the Sgr core.  
Through inspection of Sgr core distances from Table 6 and plane-plane angles from Table 7, the best way to pair the streams, assuming two dwarfs, is (a) Sgr-North and Bifurcated South; and (b) Sgr-South and Bifurcated North. The angles between these stripes pairs are 10.9${}^\circ$ and 9.6${}^\circ$, respectively.
Interestingly, these angles match the the difference in North/South orbital poles in 2MASS data 
(10.4${}^\circ \pm$ 2.6${}^\circ$) \citep{jlm05}.
The planes for Sgr North/Bifurcated South both pass over 2 kpc away from the Sgr core (-3.5 and -2.2 kpc, respectively) while Sgr South/Bifurcated North intersect (within errors) the Sgr core itself.  This seems to imply that the thick streams that have traditionally been assumed to be Sgr dwarf tidal debris may in actuality be two separate streams.

If there are more than two dwarfs (case 3), then it is difficult to say anything about the system without having a complete understanding of the stream density distribution.  
Currently, this information is unavailable, as the plane of the disk (in the direction of the Galactic center) blocks our line of sight to where these dwarfs would be discernible as separate entities.  
This description seems to add more complexity than is needed to describe the problem; however, case 1 either does not contain enough details to describe the current situation, or adds theoretical complexities (in the form of internal dynamics or dark matter distribution) that have yet to be solved satisfactorily.
From the distances to the dwarf in Table 6, it seems more likely that the Sgr dwarf is associated with the Sgr South stream and the bifurcated piece of Sgr North.

In summary, the results of the maximum likelihood optimizations and tidal debris extractions provide a set of very accurate determinations to use in construction of a better disruption model.  Not only are summary statistics available for use in determining the correctness of a simulation, but a catalog of stars fitting the density profile of the tidal stream has been generated.  This provides a brand new means in which the validity of the simulations may be tested, and will serve to constrain the models for the Galaxy.

\section{The Stellar Density and Width as a Function of Angle Along the Sgr Stream}

In addition to position and direction, the tidal streams have a width and density as a function of angle along the progenitor.  
The angle along the stream is given in the Sun-centered $\Lambda_{\odot}, B_{\odot}$ coordinate system used in \citet{lm10}, in a right-handed ($XYZ$) Galactic coordinate system.  The $\Lambda_{\odot}=0^{\circ}$ reference is the Sgr Dwarf galaxy, and increases towards the Galactic South (towards the trailing tail, or opposite the orbital direction of Sgr).  
Therefore $\Lambda_{\odot}=180^{\circ}$ is in the opposite direction from the Sgr dSph as viewed from the Sun. The leading (North) stream is traveling from $360^{\circ}$ towards $180^{\circ}$, while the trailing (South) stream is moving towards $0^{\circ}$.

The stellar density along the stream is defined as the number of Sgr tidal stream stars per degree of $\Lambda_{\odot}$ in the sky.  
To get this number, we first separated the Northern Sgr stream from the SDSS data (Section~\ref{separatingdebris}, Figure 6), then sliced this subset into 15 (the number of SDSS data stripes used) equal-width wedges, in $\Lambda_{\odot}$, $7.65\bar{3}$ degrees wide.  
We then counted the number of stars in each wedge and divided by the degree width to get counts per degree.

However, these counts do not represent the true number of F turnoff stars, as stars are missing from the data due to the SDSS detection efficiency \citep{nyetal02} and in places where the Sgr tidal stream is not fully contained within the data.  
To compensate for this, we used a Monte-Carlo integration method to infer the number of missing stars in the stripe.  
We randomly generated valid stream stars given the best-fit parameters for that stripe, as per Section 3 of \cite{cetal08}, and then tested to see if the star would have been detected by SDSS, given the F turnoff detection efficiency and stripe boundaries.  
If the star would have been detected by SDSS, it was flagged as ``detected" and added to the dataset;  otherwise, the star would be added to the dataset and not flagged.  Any star outside of the $\nu$ bounds ($-1.25^{\circ}<\nu<1.25^{\circ}$) was rejected outright to prevent overlap with adjacent stripes.  
When the number of ``detected" stars was equal to the number of stars belonging to Sgr in that stripe (Table 3), the process was stopped.  
This process resulted in a catalog of stars with a density consistent with that of the Sgr tidal stream in a given stripe, but with additional stars correcting for losses due to the stripe bounds and the detection efficiency.
After the stream in each stripe was corrected in this way, the simulated data was merged into one dataset and then sliced into $\Lambda_{\odot}$ bins and divided into wedges, as described above.

Since the great circle of $B_{\odot}=0^{\circ}$ (that is, the Sgr orbital plane given by Law \& Majewski 2010) does not intersect the SDSS data stripes at a $90^{\circ}$ angle, there will be large volumes of data missing from the $\Lambda_{\odot}$ bins located at the edge of the data. We flagged the $\Lambda_{\odot}$ stripes that are suspected to contain this bias, and plotted their densities using open points.

The SDSS South stripes used in our analysis were not contiguous, and so we approximated their densities by dividing the star counts by the length of the Sgr stream in that stripe.  
We then reconstructed the missing stars, as above, and divided by the same length to get the expected density.  
If the density of the stream is not varying rapidly near the location of the SDSS detection, then we expect our method to provide a fairly accurate description of the stream density in that data stripe.  

Although it is questionable to compare the density of stars along the stream with an n-body simulation, presumed to represent primarily dark matter particles, we nonetheless 
compare our results to the densities expected from four Sgr dSph tidal disruption models created by \citet{lm10, ljm05}.  These models are n-body simulations that best describe the Sgr tidal stream data using different gravitational potentials for the Galactic halo:  tri-axial, non-axisymmetric \citep{lm10}, spherical ($q=1$), oblate ($q<1$), and prolate($q>1$) \citep{ljm05}.  
We first selected masses from each model by ``Pcol" (which indicates the orbit on which a given 
particle became unbound from the simulated satellite) and ``Lmflag" (which is positive for leading tail 
debris, negative for trailing debris).  We selected only stars with Pcol$< 3$ (corresponding to the first 
wrap only) and $160.0 < \Lambda_{\sun} < 185.0$.  The latter condition is because, in these models, the 
leading and trailing tails each wrap more than $180^{\circ}$ around the galaxy, and so each tail will 
extend into the opposite respective Galactic hemisphere, creating a apparent overlap in angular 
density near the Galactic anti-center.  Since only the regions of the streams closest to the Sgr core
would have been characterized by our analysis, we need to select only the stream stars located in their 
respective Galactic cap.  For the tri-axial model, we selected stars with ``Lmflag"$> 0$ and $185.0 < \Lambda_{\sun} < 360.0$ to isolate the leading (Northern) tail, and ``Lmflag"$< 0$, $0.0 < \Lambda_{\sun} < 160.0$ to select only the trailing (Southern) tail.  For the axisymmetric models, cuts were made in $\Lambda_{\sun}, W$ ($z$-axis velocity) in order to separate the primary leading and trailing tails into their respective hemispheres.

Using the data from these models, we sliced the simulated streams into $1.0^{\circ}$ $\Lambda_{\odot}$ bins and counted the number of stars in each bin, thereby obtaining the stellar density along the stream.  
We plot these densities as lines in Figure 8 (top pane), multiplied by an arbitrary factor to bring them in line with the observed densities (since the number of bodies in the simulations does not match the number of stars in the actual stream).

\begin{figure}
\label{width_plot}
\figurenum{8}
\includegraphics[width=1.0\textwidth]{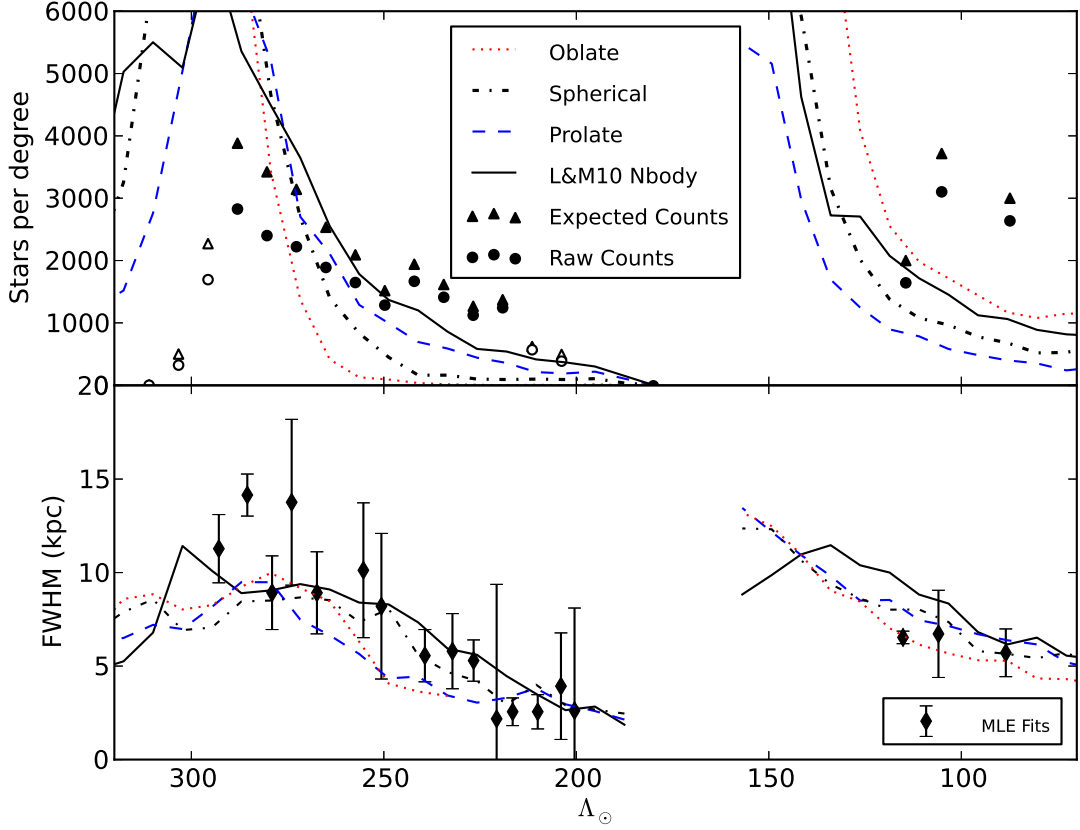}
\caption{\it Sagittarius stream density (top) and width (bottom) versus $\Lambda_{\odot}$.  In the Sgr stream density plot (top), square points represent the density as given by the number of F turnoff stars observed per $\Lambda_{\odot}$ degree in the SDSS data, while triangles show the corrected density, accounting for the SDSS detection efficiency and edge effects.  Open symbols represent points that may be missing counts due to edge effects (see text).  Typical errors are on the order of the point sizes, while maximum errors are roughly twice the point sizes.
The over-plotted lines represent the star densities from best-fitting n-body simulations in various halo potentials:  Solid black line:  triaxial halo \citep{lm10};  Red dotted line: axisymmetric oblate halo; Black dotted line: spherical halo;  Blue dotted line:  axisymmetric prolate halo.  The latter three models are from \citet{ljm05}.  All n-body densities are scaled by an arbitrary factor (1.8, 1.8, 2.0, and 2.4, respectively).
It is clear the the Southern data does not match any of the theoretical models.  
The lower plot shows the width (FWHM) of the Sgr stream versus $\Lambda_{\odot}$.  The diamond symbols and associated errorbars are the stream widths and errors as given by the maximum likelihood algorithm in this paper.  The solid line represents widths derived from the \citet{lm10} triaxial n-body model, using ``Pcol"$< 3$ and ``Lmflag" $> 0$ for leading (Northern) tail, or ``Lmflag" $< 0$ for the trailing (Southern) tail (see text).  
Note that while one could imagine a reasonable variation in the mass-to-light ratio along the stream that would match the data to the model for $\Lambda_{\odot} > 200$ (the North Galactic Cap), the data in the South ($\Lambda_{\odot} \sim 100$) do not fit the model. }
\end{figure}

It can be seen from Figure 8 that none of the models are a good match to the data. This is not 
too surprising, since the n-body models trace the path of all mass originally associated with the the 
Sagittarius dwarf, including dark matter, while our analysis only traces the luminous matter (specifically, 
F turnoff stars, which are plentiful and can be assumed to be well-mixed with other stars in the dwarf).
The results from the leading tidal tail, then, suggest that mass does not follow light exactly within the
disrupted dwarf system.  
However, we note that the density of stars along the smaller Orphan Stream can be well-fit with n-body models \citep{nyetal10}.
If one were interested in developing a function for the number of 
stars in the tidal stream for a given total mass per $\Lambda_{\sun}$ (so as to make the n-body simulations 
directly comparable to the data), the tri-axial and prolate model densities would require the least amount of correction.

The densities in the trailing tail tell an entirely different story:  only one of the three points matches any of the four models, and the remaining two points show a completely separate trend.
Interestingly, while our analysis shows the Sgr South stream to be the most consistent with the Sgr dwarf, the data from the South is not consistent with the best Sgr dwarf disruption models.
There are three possibilities for this discrepancy:  (1) The n-body models are somehow in error; (2) Our approximation for single-stripe density is incorrect; or (3) the dominant stream in the South Galactic Cap is not from Sagittarius.  Note that possibility (3) is the opposite of the conclusion reached from our fits to the plane of the tidal tails.  The plane fits favor an association between the Sgr dwarf and the dominant trailing tidal tail, as is seen in 2MASS data \citep{mswo03}.  In order to rule out possibility (2), the SDSS South data will need to be analyzed in more detail.  This is possible with the addition of SDSS Data Release 8 (DR8), which fills in large parts of the Southern Galactic cap, and will be the topic of a future paper.
If in fact the leading tidal tail is not composed of stars from the Sgr dwarf galaxy, then it would be natural for the n-body simulations to be in error, since they were optimized using the leading tidal tail stars.

The bottom pane of Figure 8 displays the stream FWHM (full-width at half-maximum) as a function of $\Lambda_{\odot}$, as well as the stream widths from \citet{lm10, ljm05}.  
Our FWHMs and errorbars are derived by multiplying the $\sigma$ stream width values and errors (Table 3) by 2.35. 
We sliced the n-body model stars into wedges of size $\Lambda_{\odot}=7.65\bar{3}$ (as above).  In each $\Lambda_{\odot}$ wedge, we assumed that each star had a $\Lambda_{\odot}$ equal to the middle value of the wedge, thereby creating a 2D plane of data in $R,B_{\odot}$.  
We then defined a 2D Cartesian system in this plane, with $\hat{x}$ being along the line-of-sight, and $\hat{y}$ being the perpendicular direction, such that $x=R\cos B_{\odot}$ and $y=R\sin B_{\odot}$.  
The $x$ and $y$ standard deviations were then determined for each wedge, and the geometric mean of these values then gives the cylindrical stream width:  $\sigma_{xy} = \sqrt[2]{\sigma_x \cdot \sigma_y}$.   
We note that the n-body stream was not cylindrical - the $x$ (or radial) standard deviations were almost always larger than the $y$ standard deviations.  
The stream profile, then, is more elliptical in the simulations than our assumed circular profile.  
Even if this is the case in the actual Sgr tidal stream, the assumed-cylindrical $\sigma$ fit from our algorithm should be analogous to $\sigma_{xy}$, and therefore it is valid to compare the two.

We see from Figure 8 that all four of the n-body models follow roughly the same width trend in the leading tidal tail as our best-fit widths.  The stream widths in the South are also consistent with the model, although the actual data appears to be a bit flatter than the model.  
Our stream widths are raw parameters from the data fits, so we expect these to be accurate.
Again, more information about the South is required before an accurate comparison can be made, but at this point, the widths appear to be in good agreement with both the \citet{lm10} model and the spherical potential model from \citet{ljm05}, for both the dominant leading and dominant trailing tidal tails.  
The oblate and prolate potentials produce width profiles that are a worse fit to the data, but still within the given errors.

\section{Fit Parameters for the Spheroid Star Density}

Since spheroid parameters were measured for each SDSS stripe individually, we have 18 measurements of these parameters.  
If the Hernquist profile is a good fit to the smooth stellar density in the halo (which may not be true; see below), then the parameters should be similar for each of the stripes.

The Hernquist $q,r_0$ parameter fits to the SDSS stripe data are presented in Table 8.  We find that the stellar spheroid is clearly oblate, as the $q$ value for every stripe is less than $1.0$ (spherical).  
The Northern Galactic cap has very consistent $q$ values, with a mean value of $0.53$ and a standard deviation in the fit values of 0.06.  There is no apparent trend in $q$ with stripe number in the Northern cap, and the errors in the fits are larger than the average difference in $q$ fits, and so we feel that the halo flattening has been well-fit in the North.

The South Galactic Cap exhibits a broader range of $q$ fit values, despite consisting of only three stripes.  
The Southern cap contains an apparent decrease in $q$ with increasing angle away from the Sgr core, although the errors in these values places them within $2\sigma$ of the Northern values, so this may not be significant.
The average and standard deviation of the Southern $q$ values are 0.48 and 0.12, respectively.  If the Southern and Northern cap $q$'s are included together, the average value of $q$ becomes $0.52$, with a standard deviation in the fit values of 0.08.

The Hernquist profile scale length, $r_0$, varies widely in our parameter fits, varying between $1.84 < r_0 < 11.55$ kpc in the Northern cap, and between $1.84 < r_0 < 25.95$ kpc over all stripes.  
The model errors in the $r_0$ fits are also large (on the order of $\sim$50\% in the North).  This is partially due to fact that the algorithm is not as sensitive to changes in $r_0$ as it is for the other parameters \citep{cetal08}.  
There is no consistent trend with angle along the stream in the North Galactic Cap, and so the variations are most likely due to random errors in the data, or the lack of sensitivity of the algorithm.  
Within errors, the Northern $r_0$ values are consistent, with an average value of 6.73 kpc and a standard deviation of 2.51 kpc.  

The three Southern stripes appear to have an increasing trend in $r_0$ with angle away from the Sgr core. However, the large errors on these values imply that this correlation is a weak one.  
That the Southern values for $r_0$ are significantly larger than the Northern values (even after accounting for errors) implies that there is a difference in the stellar spheroid between the North and the South.  
This could be due to the presence of the Hercules-Aquila cloud, which is present in SDSS South but not accounted for in our analysis, or this may be additional evidence that the smooth stellar spheroid is non-uniform, or this could indicate that a Hernquist profile is not a good fit to the smooth component.
The mean and standard deviation of $r_0$ in the Southern stripes is 20.71 and 3.89 kpc, respectively.  If we analyze the Northern and Southern stripes together, the average of $r_0$ becomes 9.06 kpc, and the standard deviation rises to 5.91 kpc.

It is quite possible that the Hernquist profile is not a good descriptor of the Galactic halo stellar density.  
The insensitivity of the algorithm to the fit parameter $r_0$ is possible evidence of this.  
More problematic, however, is the fact that the ``garbage collection" stream (Section 6) always converged to a set of parameters that would collect stars at the edge of each data set, implying that the Hernquist profile by itself did not adequately describe the smooth stellar halo.
We note that the algorithm did not have problems when fitting the Hernquist profile to simulated data \citep{cetal08}, which strongly implies that the real data is not well-fit by the Hernquist profile.
Since no ``garbage collection" stream was fit to stripes in the South, it is possible that the discrepant $q$ and $r_0$ values found there are due to the smooth component having to compensate for the density that would have been accounted for by that stream.  
The additional unknown, here, is whether the selected F turnoff stars, particularly those near the magnitude limit of the survey, are good tracers of the spheroid density.  
We are working on correcting for selection effects due to increasing color errors near the survey limit, and will present those result in a future publication.
Note that \citet{cetal08} showed that the stream parameters were robust to improperly specified smooth stellar halo component models, so even if the Hernquist profile is eventually found to be a bad choice, the stream parameters should still be reliable.  

We find several similarities in our halo fits with the findings of \citet{jetal08}.  
In that paper, the authors used an earlier release of SDSS and photometric parallax methods to fit disk and power-law halo profiles to Milky Way star densities within 30 kpc of the Sun.  
Although the power-law and Hernquist profiles are not directly comparable, the \citet{jetal08} 
analysis also found a clearly oblate stellar halo, with flattening parameter 
$q_h=0.64$, but consistent with $0.5 < q_h < 0.8$, which is in good agreement with our flattening value of 
$q = 0.53$.
\citet{jetal08} also found that their density profile had difficulty matching the densities at the edge of 
their data, a problem which manifested itself in our analysis through the ``garbage collection" stream.
We agree with their analysis that a single density function may not be adequate for describing
the Milky Way stellar halo, and we intend to explore multiple halo density profiles (including double-halo 
profiles) in a future paper.

In summary, the smooth component fits to the stellar halo are very consistent in the North Galactic Cap, using a Hernquist profile with $q = 0.53$, and $r_0 = 6.73$ kpc.  The data in the South is not as consistent, however, due to either additional over-densities (such as Hercules-Aquila), deficiencies in the Hernquist profile, or the fitting technique.

\section{Verification Using MilkyWay@home}\label{milkywayathometechniques}

We used MilkyWay@Home to provide a secondary verification of our best-fit parameters. MilkyWay@home uses the Berkeley Open Infrastructure for Network Computing (BOINC) \citep{anderson_boinc_2005}, which provides a software infrastructure for a distributed, volunteer computing network. MilkyWay@Home has around $30,000$ active (at any given time) users who provide 0.5 PetaFLOPS ($10^{15}$ floating-point operations per second) of computing power\footnote[1]{http://boincstats.com/}. By necessity and design, it is a highly asynchronous platform running modern numerical optimization techniques on users’ spare computer processing power, as well as graphical processing unit (GPU) hardware, with GPUs providing significant performance improvements \citep{desell_ppam_milkyway_gpu_2009}.

When we first began work towards this paper, MilkyWay@home was still in the early phases of its development, and so our initial local optimization methods were performed on a single-rack IBM BlueGene/L (the Rensselaer Polytechnic Institute's SUR BlueGene) consisting of 512 dual core processors. Once the Milkyway@home platform was mature, we began to run evolutionary algorithms for optimization on MilkyWay@home alongside the BlueGene/L.  
We were unable to use the previously developed conjugate gradient descent optimization because Milkyway@home operates in an asynchronous, heterogeneous computing environment.
As each platform was using different search methods, but identical likelihood calculations, this allowed us to compare the performance and accuracy of different computing systems and optimization methods.

MilkyWay@home uses the Toolkit for Asynchronous Optimization (TAO) \footnote{The TAO source code be found at: https://github.com/travisdesell/tao} to optimize the search parameters over a wide range of possible parameter sets.
Currently, Milkyway@home runs the differential evolution, genetic search, and particle swarm optimization methods \citep{desell-phd-2009}.
Although each of these methods returned similar results, we chose to use particle swarm optimization as the standard for verifying our results, as it typically provided the most accurate results using the least amount of computing power.

With MilkyWay@home, we were able to get accurate results for each stripe in 1-2 weeks, with 4-6 stripes running simultaneously. For comparison, with relevant factors taken into account, the BlueGene would take 2-4 weeks per single stripe. Since the numerical optimization methods used by MilkyWay@home were global optimization techniques (that is, they used randomized evolutionary algorithms which are much less likely to be trapped in a local optimum) they did not require ``by eye" starting points and repeated runs. However, stripe-to-stripe continuity conditions were difficult to enforce on MilkyWay@home. While the local search method used on the BlueGene/L (gradient descent) allowed us, to start from an initial set of parameters that are continuous with neighboring stripes and find the nearest minimum, the global optimization methods explore the entire parameter space to find the parameter set with the best likelihood, even if it does not make physical sense in light of additional information from adjacent stripes. To compensate for this shortcoming, we enforced bounds on the fit parameters that we shrunk with each successive fit. Initially, all fits were unbound, and in the final iteration all fit parameters were bound at $\pm$20\% of the average parameters of the adjacent stripes.  This constraint forced the searches to look for parameters that were consistent with adjacent stripes, but was still allowed to search a range of parameters in detail.

We present the results of the MilkyWay@home fits to stripes 10-23 in Table 9. Additionally, the ``$N_{\sigma}$" values are presented; these are the number of standard deviations (from Table 3) that Milkway@home differs from the BlueGene results. The two sets of best-fit parameters are very much in agreement with each other; most of the differences are less then 1-$\sigma$ apart. The only stripes with $N_{\sigma} >$ 2 are stripes 11 and 23. For these stripes, the only outlying parameter is $r_0$, to which we have already shown the algorithm is not very sensitive. 

We have shown that MilkyWay@home is a fast, powerful and accurate platform for analyzing astronomical data \citep{desell-validating-volunteers-2010,desell-analysis-massive-eas-2010}. Additionally, Milkyway@home succeeded in producing the same or better results than the BlueGene searches, with virtually no artificial constraints.  In the future, we plan to use it to extend our analysis of the Galactic halo to the Galactic South (using data from SDSS DR8) and to fit the additional structure in the Northern Cap (Sgr Bifurcation, Virgo). Additionally, n-body searches are currently being implemented on the system, to extend future analysis to include orbit fits and to probe the distribution of the Dark Matter halo \citep{desell-milkyway-nbody-2011}.

\section{Conclusions}\label{conclusions}

In this paper we fit the stellar density of the Sgr dwarf tidal stream in data from 18 SDSS stripes, using a maximum likelihood technique.  The major conclusions are as follows:

(1) We successfully applied and extended to multiple streams and to using data from adjacent stripes, the Maximum-Likelihood Evaluator developed in \citet{cetal08} to the data available in SDSS DR7, a technique now called statistical photometric parallax.  We found a smooth, continuous fit for the Sgr Tidal Stream, while avoiding contamination from additional substructures in the Northern Galactic cap, as demonstrated in Figures 5-7.

(2) We present a consistent set of parameters that describe the position, width and density along the dominant Sgr leading and trailing tidal tails.  These parameters and their associated model errors are presented in Table 3.  The positions in common coordinate systems are tabulated in Table 4.

(3) We include a catalog of stars consistent with the density profile of the Sgr tidal stream, with the data given in Table 5 and displayed in Figure 6.  While this catalog does not represent true membership in Sgr, it provides a set of stars with a profile similar to Sgr that can be used for constraining simulations and theoretical models.

(4) We compare the width and stellar densities of the dominant Sgr tidal tails, as a function of angle along the stream, with popular models of Sgr dwarf galaxy disruption.  
Though the comparison of n-body calculations with the spatial density of F turnoff stars is questionable, the data are reasonably consistent with popular models in the North Galactic Cap, but largely inconsistent in the South Galactic Cap.
Since only three stripes in the South were analyzed these results should be verified with additional data in the South.

(5) By fitting planes to the dominant Sgr streams (Table 6), and to the bifurcated pieces as described in other works, we we conclude that it is difficult to reconcile the large discrepancy between the thick stream planes ($15.6\pm0.1^{\circ}$) and the presumed bifurcated pieces ($3.9\pm2.2^{\circ}$) if one assumes the Sgr dwarf galaxy is the sole progenitor.  If one assumes the tidal debris comes from the disruption of two dwarf galaxies, the plane analysis favors an association of the dominant trailing stream and the ``bifurcated" Northern stream with the Sgr dwarf.  This is the opposite conclusion of the stream width analysis, which favors an association of the dominant Northern tidal tail and the Sgr dwarf.

(6) We find that the stellar halo in the North Galactic Cap, assuming a Hernquist density profile, can be described by $q=0.53$ and $r_0 = 6.73$.  The dispersions in these measurements are $0.06$ and $2.51$, respectively.  We can not make strong conclusions about the South Galactic Cap, although the stellar halo is almost certainly oblate ($q < 1$).  We also find that the Hernquist profile may not be a good fit to the smooth stellar spheroid, or that selection effects near the survey limit are problematic.

(7) We verify our parameter fit results with the powerful Milkyway@home computing platform.  In doing so, we show that  Milkyway@home is mature and capable of producing results consistent with other, less automated and more time-consuming methods.  

In summary, we provide the best measurements to date of the properties of the dominant Northern and Southern tidal streams thought to be associated with the Sgr dwarf galaxy.  As in the past, a definitive model for the creation of these tidal streams remains elusive.

\acknowledgments

We would like to thank our referee, David Law, for comments which improved the quality of this paper.
Portions of this text have been copied or adapted from the PhD Thesis of Nathan Cole \citep{colethesis}.
We would also like to thank the over 140,000 participants who have contributed to Milkyway@home over the past several years.  Most participants have volunteered computing time to our project, although many participants have contributed financial resources or equipment, several participants have contributed to error-checking and troubleshooting, and a few participants have assisted in developing the project software.  We would like to acknowledge and thank David Glogau and The Marvin Clan for their significant donations toward the project.  We would also like to thank the Dudley Observatory for their assistance.  This publication is based upon work supported by the National Science Foundation under Grant No. AST 10-09670.

We use data from the Sloan Digital Sky Survey.
Funding for the SDSS and SDSS-II has been provided by the Alfred P. Sloan Foundation, the Participating 
Institutions, the National Science Foundation, the U.S. Department of Energy, the National Aeronautics and 
Space Administration, the Japanese Monbukagakusho, the Max Planck Society, and the Higher Education Funding 
Council for England. The SDSS Web Site is http://www.sdss.org/.

The SDSS is managed by the Astrophysical Research Consortium for the Participating Institutions. The Participating 
Institutions are the American Museum of Natural History, Astrophysical Institute Potsdam, University of Basel, 
University of Cambridge, Case Western Reserve University, University of Chicago, Drexel University, Fermilab,  
the Institute for Advanced Study, the Japan Participation Group, Johns Hopkins University, the Joint Institute 
for Nuclear Astrophysics, the Kavli Institute for Particle Astrophysics and Cosmology, the Korean Scientist 
Group, the Chinese Academy of Sciences (LAMOST), Los Alamos National Laboratory, the Max-Planck-Institute 
for Astronomy (MPIA), the Max-Planck-Institute for Astrophysics (MPA), New Mexico State University, Ohio 
State University, University of Pittsburgh, University of Portsmouth, Princeton University, the United States 
Naval Observatory, and the University of Washington.

\clearpage

\appendix
\section{APPENDIX A:  Plane-Fitting Methods and New In-Plane Coordinates}\label{app-A}

\subsection{Fitting a Plane to Galactic Data}

In this section, we describe the methods used to fit planes to stream detections in section 7.1, and define a new set of $\Lambda_{\rm new}, B_{\rm new}$ coordinates to reflect the new Sgr orbital plane.  
This discussion was omitted from the main body of the paper as we deemed that it would clutter the data analysis sections, but we still wished to include this appendix to provide a more complete discussion of our error analysis, and to aid future researchers in similar pursuits.  We will assume a plane equation of:

\begin{equation}\label{plane}
ax + by + cz + d = 0.
\end{equation}

When fitting an orbital plane to data, it is preferable to fit the direction of the normal vector in polar coordinates, rather than in Cartesian $xyz$.  
Since we can assume the normal is a unit vector, this reduces the number of fitting parameters from 4 to 3 ($[(a,b,c,d) \to (\theta,\phi,d)]$), where $\phi$ is the angle from the galactic $z$-axis, and $\theta$ is the angle from Galactic $x$-axis, measured counter-clockwise (right-handed) about the $z$-axis.  
The coordinate systems are related by the following: 
\begin{equation}
\begin{array}{lll}
a=\cos\theta \sin\phi \\
b=\sin\theta \sin\phi \\
c=\cos\phi
\end{array}
\end{equation}
and errors are therefore given by:
\begin{equation}
\begin{array}{lll}
\delta a = -\sin\theta \sin\phi \delta\theta + \cos\theta \cos\phi \delta\phi \\
\delta b = \cos\theta \sin\phi \delta\theta + \sin\theta \cos\phi \delta\phi \\
\delta c = -\sin\phi \delta\phi
\end{array}
\end{equation}
and with $\delta d = \delta d$, of course.  The $\chi^2$ of this method is simple as well:
\begin{equation}
\chi^2 = \sum_n \frac{D_{n}^2}{\sigma_{D_n}^{2}}
\end{equation}
where $D_n$ is the distance from the plane to point $n$, and $\sigma_{D} = a \sigma_x + b \sigma_y + c \sigma_z$ for point $n$.

Since angles are cyclical, the $\theta, \phi$ $\chi^2$ surface will repeat itself ad-infinitum, and artificial bounds on the solution space are unneeded.  
The derived parameters $(a,b,c)$ will be automatically normalized by this method.
A simple gradient search method will perform very well in this situation, as the search will always be pointed at one of the many equivalent minimums, and will converge quickly.  
More global methods such as Monte-Carlo Markov Chains (MCMC) will fail here, however, as the existence of multiple equivalent minimums will confuse the statistical nature of this type of search.

Errors in the fit parameters were found via the Hessian method.  First, a Hessian matrix ($\mathcal{H}$) of size $m \times m$ of 2nd-order partial derivatives is created, where $m=$ number of fit parameters.  After inverting the Hessian, the diagonal elements become equivalent to half of the variances ($\sigma^2$) in the parameter fits.  Therefore, the error in parameter $i$ is equal to:
\begin{equation}
\sigma_i = \sqrt{2 \mathcal{H}^{-1}_{i,i}}.
\end{equation}

After the best-fit plane normal is determined, we standardize our results by defining the parameter $c$ to always be positive (multiplying Equation~\ref{plane} by -1 if necessary), causing the plane normal to always point as much in the direction of the Galactic $z$-axis as possible.  In the rare case that $c=0$, then we define $b$ to be the positive parameter.

\subsection{A Standard and Simple Plane Coordinate System}

We now define two simple-to-use in-plane coordinate transformation systems, one Cartesian, the other spherical, that do not require the careful use of Euler angles.  We begin with a unit vector normal to the plane, $\hat{n}=(a,b,c)$, and a point in Sun-centered Galactic Cartesian coordinates, $\vec{P_0}=(X_0,Y_0,Z_0)$, that will become the rough direction of the new $x$-axis (We use the convention with $X$ positive in the direction of the Galactic Center, so that the coordinate system is right-handed). 
Normalizing $\vec{P_0}$ gives:  $\hat{p} = \frac{\vec{P_0}}{|\vec{P_0}|} = (x_0, y_0, z_0)$ .
We can then define the new in-plane (primed) axes using vector cross-products:
\begin{displaymath}
\begin{array}{lll}
\hat{z}' = \hat{n} \\
\hat{y}' = \hat{z}' \times \hat{p} \\
\hat{x}' = \hat{y}' \times \hat{z}' = (\hat{n} \times \hat{p}) \times \hat{n}.
\end{array}
\end{displaymath}
Note that $\hat{x}'$ does not necessarily point exactly in the direction of $\hat{p}$;  this is intentional, so that one is not required to carefully select a $\hat{p}$ that is perpendicular to $\hat{n}$.
The in-plane Cartesian coordinates ($\vec{R}'$) for a point $\vec{R}=(x,y,z)$ in Galactic coordinates is, using vector dot-products:
\begin{equation}
\vec{R}' = (x \cdot \hat{x}', y \cdot \hat{y}', z \cdot \hat{z}') = (x',y',z')
\end{equation}
We can then define in-plane longitude and latitude as $\Lambda, B$:
\begin{displaymath}
\begin{array}{ll}
\Lambda = \rm{arctan2}(y', x') \\
B = \arcsin \left(\frac{z'}{\sqrt{x'^2 + y'^2}} \right)
\end{array}
\end{displaymath}
where $0^{\circ} \le \Lambda \le 360^{\circ}$ is the in-plane longitude, and $-90^{\circ} \le B \le 90^{\circ}$ is the in-plane latitude.  
By defining $c$ in the plane normal to be positive (previous subsection), we set $z'$ to increase in the direction of Galactic $z$, and $\Lambda$ increases in the direction of increasing Galactic $l$, while $B$ increases in the direction of increasing Galactic $b$.  

This system is superior to Euler angle rotations, which occasionally cause confusion when it comes to the direction and order of the necessary rotations.  Additionally, this coordinate transformation system is computationally less expensive than the Euler rotation method, since it requires fewer calls to trigonometric functions per data point, especially when working in in-plane Cartesian coordinates.
Since this definition of in-plane coordinate transforms depends on two normalized vectors only (the plane normal and rough direction of eventual x-axis), and the conversion requires the use of unambiguous vector dot- and cross-products, these coordinates are easy to define, derive, use, and can be applied to currently-existing planar coordinate systems.

We present our Sgr tidal stream best-fit center points in $\Lambda_{new}, B_{new}$ in Table 4, using a $\hat{n}=(-0.199, 0.935, 0.293)$ from the Sgr North plane fit in Section 7.1, and $\vec{P_0}=(23.156, 2.270, -5.887)$, the Sun-centered location of the Sgr dSph \citep{mswo03}.  To apply these transforms to the commonly-used $\Lambda_{\odot}, B_{\odot}$ system, $\hat{n}=(-0.064, 0.970, 0.233)$ and $\vec{P_0}=(23.156, 2.270, -5.887)$ should be used.

\clearpage

\clearpage
\begin{deluxetable}{lccccccc}
\tabletypesize{\scriptsize}
\tablecolumns{8}
\footnotesize
\tablecaption{SDSS Stripe Data Characteristics\tablenotemark{1} \label{DataChar}}
\tablewidth{0pt}
\tablehead{
\colhead{Stripe} & \colhead{$\mu_{min}$} & \colhead{$\mu_{max}$} & \colhead{$g_{0min}$} & \colhead{$g_{0max}$} & \colhead{\# stars} & \colhead{\# streams fit} & \colhead{\# cuts}}

\startdata
9	& 170$^{\circ}$ & 235$^{\circ}$ & 16.0 & 23.5 & 95,435  & 3 & 0 \\
10	& 165$^{\circ}$ & 227$^{\circ}$ & 16.0 & 23.5 & 97,939  & 3 & 0 \\
11	& 150$^{\circ}$ & 229$^{\circ}$ & 16.0 & 23.0 & 97,434  & 3 & 0 \\
12	& 135$^{\circ}$ & 235$^{\circ}$ & 16.0 & 23.0 & 120,612 & 3 & 0 \\
13	& 135$^{\circ}$ & 235$^{\circ}$ & 16.0 & 22.5 & 118,836 & 3 & 0 \\
14	& 135$^{\circ}$ & 235$^{\circ}$ & 16.0 & 22.5 & 102,599 & 3 & 0 \\
15	& 135$^{\circ}$ & 240$^{\circ}$ & 16.0 & 22.5 & 108,460 & 3 & 0 \\
16	& 135$^{\circ}$ & 240$^{\circ}$ & 16.0 & 22.5 & 107,033 & 3 & 0 \\
17	& 135$^{\circ}$ & 235$^{\circ}$ & 16.0 & 22.5 & 91,626  & 3 & 2 \\
18	& 135$^{\circ}$ & 240$^{\circ}$ & 16.0 & 22.5 & 95,462  & 3 & 1 \\
19	& 135$^{\circ}$ & 230$^{\circ}$ & 16.0 & 22.5 & 84,046  & 3 & 0 \\
20	& 133$^{\circ}$ & 249$^{\circ}$ & 16.0 & 22.5 & 105,909 & 3 & 0 \\
21	& 133$^{\circ}$ & 210$^{\circ}$ & 16.0 & 22.5 & 60,503  & 3 & 0 \\
22	& 131$^{\circ}$ & 225$^{\circ}$ & 16.0 & 22.5 & 66,200  & 3 & 2 \\
23	& 133$^{\circ}$ & 230$^{\circ}$ & 16.0 & 22.5 & 65,335  & 3 & 1 \\
\tableline
79	& 311$^{\circ}$ & 416$^{\circ}$ & 16.0 & 22.5 & 92,789  & 1 & 0 \\
82	& 310$^{\circ}$ & 419$^{\circ}$ & 16.0 & 22.5 & 115,884 & 1 & 0 \\
86	& 310$^{\circ}$ & 420$^{\circ}$ & 16.0 & 22.5 & 111,642 & 1 & 1 \\
\enddata
\tablenotetext{1}{In all cases, the $\nu$ range is $-1.25 < \nu < 1.25 $}
\end{deluxetable}

\begin{deluxetable}{lccccl}
\tabletypesize{\scriptsize}
\tablecolumns{6}
\footnotesize
\tablecaption{Excluded Sections of SDSS Stripes \tablenotemark{1} \label{CutChar}}
\tablewidth{0pt}
\tablehead{
\colhead{Stripe} & \colhead{$\mu_{min}$} & \colhead{$\mu_{max}$} & \colhead{$\nu_{min}$} & \colhead{$\nu_{max}$} & \colhead{Reason}}
\startdata
17 & 182.4$^{\circ}$ & 183.0$^{\circ}$ &  0.9$^{\circ}$  & 1.25$^{\circ}$ & Globular Cluster \\
17 & 197.0$^{\circ}$ & 199.0$^{\circ}$ &  0.4$^{\circ}$  & 1.25$^{\circ}$ & Globular Cluster \\
18 & 197.0$^{\circ}$ & 198.0$^{\circ}$ & -1.25$^{\circ}$ & -1.0$^{\circ}$ & Globular Cluster \\
22 & 207.0$^{\circ}$ & 209.0$^{\circ}$ &  0.8$^{\circ}$  & 1.25$^{\circ}$ & Globular Cluster \\
22 & 202.0$^{\circ}$ & 204.0$^{\circ}$ & -0.5$^{\circ}$  & 0.4$^{\circ}$  & Globular Cluster \\
23 & 207.0$^{\circ}$ & 209.0$^{\circ}$ & -1.25$^{\circ}$ & -0.8$^{\circ}$ & Globular Cluster\\
86 &  21.6$^{\circ}$ &  22.3$^{\circ}$ & -1.25$^{\circ}$ & 1.25$^{\circ}$ & Data Artifact\\
\enddata
\tablenotetext{1}{In all cases, the $g_0$ range spans the entire magnitude range of the stripe, 
$16.0 < g_0 < 22.5 $}
\end{deluxetable}

\begin{deluxetable}{lccccccr}
\tabletypesize{\scriptsize}
\tablewidth{0pt}
\tablecolumns{8}
\tablecaption{ Sagittarius Stream Parameters \label{ sgrtable }}
\tablehead{
\colhead{Stripe} & \colhead{$\epsilon$} & \colhead{$\mu ({}^{\circ})$} & \colhead{$R$ (kpc)} & \colhead{$\theta (rad)$} & \colhead{$\phi (rad)$} & \colhead{$\sigma$} & \colhead{\# Stars}}
\startdata
9  & -0.4 $\pm$ 0.6 & 220.0 $\pm$ 3.0  & 43.0 $\pm$ 3.0  & 1.4 $\pm$ 0.4 & -1.8 $\pm$ 0.4 & 4.8 $\pm$ 0.8 & 19,386\\
10 & -1.4 $\pm$ 0.4 & 210.0 $\pm$ 20.0 & 44.0 $\pm$ 5.0  & 1.1 $\pm$ 0.5 &  3.0 $\pm$ 2.0 & 6.0 $\pm$ 0.5 & 18,734\\
11 & -1.3 $\pm$ 0.3 & 206.0 $\pm$ 1.0  & 41.0 $\pm$ 13.0 & 1.2 $\pm$ 1.3 & -2.9 $\pm$ 0.5 & 3.8 $\pm$ 0.8 & 11,095\\
12 & -1.3 $\pm$ 0.4 & 202.0 $\pm$ 5.0  & 41.0 $\pm$ 6.0  & 1.3 $\pm$ 0.5 &  3.1 $\pm$ 0.3 & 5.8 $\pm$ 1.9 & 17,044\\
13 & -1.0 $\pm$ 0.4 & 200.0 $\pm$ 12.0 & 36.0 $\pm$ 5.0  & 1.5 $\pm$ 0.2 & -3.0 $\pm$ 1.0 & 3.8 $\pm$ 0.9 & 18,736\\
14 & -1.4 $\pm$ 0.6 & 184.0 $\pm$ 7.0  & 28.0 $\pm$ 9.0  & 1.8 $\pm$ 0.8 &  2.9 $\pm$ 0.3 & 4.0 $\pm$ 1.0 & 15,409\\
15 & -1.7 $\pm$ 0.6 & 180.0 $\pm$ 10.0 & 27.0 $\pm$ 8.0  & 1.9 $\pm$ 0.7 &  3.0 $\pm$ 0.6 & 3.0 $\pm$ 2.0 & 12,519\\
16 & -1.5 $\pm$ 0.6 & 169.0 $\pm$ 7.0  & 28.0 $\pm$ 6.0  & 2.3 $\pm$ 0.7 &  2.9 $\pm$ 0.5 & 2.4 $\pm$ 0.6 & 12,248\\
17 & -1.6 $\pm$ 0.7 & 162.0 $\pm$ 7.0  & 26.0 $\pm$ 6.0  & 2.0 $\pm$ 0.8 &  2.9 $\pm$ 0.3 & 2.5 $\pm$ 0.8 &  8,853\\
18 & -2.0 $\pm$ 0.8 & 157.0 $\pm$ 7.0  & 25.0 $\pm$ 7.0  & 2.1 $\pm$ 0.7 &  3.0 $\pm$ 1.0 & 2.2 $\pm$ 0.5 &  7,328\\
19 & -1.9 $\pm$ 0.4 & 151.0 $\pm$ 7.0  & 23.0 $\pm$ 7.0  & 2.4 $\pm$ 1.6 &  3.0 $\pm$ 2.0 & 0.9 $\pm$ 3.0 &  5,479\\
20 & -2.8 $\pm$ 0.5 & 148.0 $\pm$ 7.0  & 23.0 $\pm$ 6.0  & 2.3 $\pm$ 0.3 &  3.0 $\pm$ 0.7 & 1.1 $\pm$ 0.3 &  4,450\\
21 & -2.1 $\pm$ 0.3 & 140.0 $\pm$ 15.0 & 21.0 $\pm$ 8.0  & 2.6 $\pm$ 0.9 &  3.0 $\pm$ 2.0 & 1.1 $\pm$ 0.4 &  3,486\\
22 & -2.0 $\pm$ 1.0 & 140.0 $\pm$ 14.0 & 18.0 $\pm$ 8.0  & 2.4 $\pm$ 1.0 &  1.9 $\pm$ 2.5 & 2.0 $\pm$ 1.0 &  2,425\\
23 & -4.0 $\pm$ 1.0 & 130.0 $\pm$ 16.0 & 17.0 $\pm$ 18.0 & 2.7 $\pm$ 1.6 &  1.4 $\pm$ 1.6 & 1.0 $\pm$ 2.0 &    971\\
\tableline
79 & -2.4 $\pm$ 0.4 & 38.0 $\pm$ 5.0 & 30.0 $\pm$ 5.0 & 2.2 $\pm$ 0.5 & 0.3 $\pm$ 0.6 & 2.8 $\pm$ 0.1 &   9,511\\
82 & -1.8 $\pm$ 0.3 & 31.0 $\pm$ 3.0 & 29.0 $\pm$ 3.0 & 1.7 $\pm$ 0.5 & 0.0 $\pm$ 1.0 & 3.0 $\pm$ 1.0 &  16,119\\
86 & -1.7 $\pm$ 0.4 & 16.0 $\pm$ 5.0 & 26.0 $\pm$ 3.0 & 1.4 $\pm$ 0.1 & 0.1 $\pm$ 0.2 & 2.4 $\pm$ 0.5 &  16,603\\
\enddata
\end{deluxetable}

\begin{deluxetable}{lcccccccccc}
\tabletypesize{\scriptsize}
\tablewidth{0pt}
\tablecolumns{11}
\tablecaption{ Sagittarius Stream Centers \label{ coordtable }}
\tablehead{
\colhead{Stripe} & \colhead{$X_{\sun} (kpc)$} & \colhead{$Y_{\sun} (kpc)$} & \colhead{$Z_{\sun} (kpc)$} & \colhead{$l$} & \colhead{$b$} & \colhead{$\Lambda_{\odot}$} & \colhead{$B_{\odot}$} & \colhead{$\Lambda_{new}$\tablenotemark{1}} & \colhead{$B_{new}$\tablenotemark{1}} & \colhead{$R_{\sun} (kpc)$} }
\startdata
9  &  18.3 & -5.3 & 33.8 & 348.8 & 51.0 &  -66.6 &-1.3 &  -74.6 & -0.5 & 43.4 \\
10 &  14.4 & -7.9 & 37.3 & 340.9 & 57.0 &  -74.0 & 0.6 &  -82.2 & -1.4 & 44.5 \\
11 &   8.7 & -8.8 & 36.3 & 332.9 & 62.0 &  -80.3 & 1.6 &  -88.6 & -1.4 & 41.1 \\
12 &   5.2 & -9.3 & 37.1 & 325.9 & 66.0 &  -85.4 & 1.7 &  -93.6 & -0.7 & 40.7 \\
13 &  -0.2 & -9.2 & 34.0 & 312.1 & 70.0 &  -91.9 & 2.4 & -100.1 & -0.4 & 36.2 \\
14 &  -7.7 & -9.3 & 26.9 & 275.1 & 70.9 & -103.7 & 5.6 & -112.3 & -1.9 & 28.5 \\
15 &  -9.9 & -8.8 & 25.8 & 260.8 & 71.0 & -108.3 & 5.1 & -116.8 & -0.8 & 27.3 \\
16 & -15.2 &-10.3 & 24.9 & 236.9 & 63.7 & -119.7 & 7.8 & -128.3 & -2.1 & 27.8 \\
17 & -17.7 & -9.9 & 22.0 & 226.9 & 58.5 & -126.8 & 8.6 & -135.4 & -2.1 & 25.8 \\
18 & -19.7 & -9.6 & 20.6 & 220.5 & 54.3 & -132.3 & 8.6 & -140.8 & -1.6 & 25.4 \\
19 & -21.0 & -8.9 & 17.8 & 215.5 & 49.3 & -138.3 & 9.1 & -146.7 & -1.6 & 23.5 \\
20 & -21.9 & -8.4 & 16.3 & 212.0 & 45.9 & -142.5 & 8.8 & -150.9 & -1.0 & 22.7 \\
21 & -22.7 & -8.0 & 13.5 & 209.2 & 39.7 & -149.0 & 10.0 & -157.4 & -1.8 & 21.2 \\
22 & -21.9 & -6.9 & 10.1 & 207.2 & 33.9 & -155.0 & 11.1 & -163.3 & -2.6 & 18.1 \\
23 & -21.4 & -6.2 &  8.5 & 205.7 & 30.7 & -158.5 & 11.1 & -166.8 & -2.6 & 16.7 \\
\tableline
79 & -27.7 & 5.8 & -22.6 & 163.3 & -48.4 & 114.4 & -2.9 & 105.8 & 5.0 & 30.2 \\
82 & -23.2 & 5.6 & -24.7 & 159.2 & -57.6 & 105.1 & -1.2 &  96.8 & 1.8 & 29.2 \\
86 & -14.1 & 5.6 & -24.8 & 134.8 & -72.3 & 87.4  & -0.0 &  79.5 & -2.1 & 26.1 \\
\enddata
\tablenotetext{1}{Defined in Appendix~\ref{app-A}}
\end{deluxetable}

\begin{deluxetable}{ccc}
\tabletypesize{\scriptsize}
\tablewidth{0pt}
\tablecolumns{3}
\tablecaption{ Sagittarius Star Catalog\tablenotemark{1} \label{ sgr_stub }}
\tablehead{
\colhead{$l ({}^{\circ})$} & \colhead{$b ({}^{\circ})$} & \colhead{$g$ (mag)} }
\startdata
217.15623 & 38.10350 & 20.5 \\
217.26543 & 38.42290 & 22.2 \\
217.49472 & 38.91518 & 21.4 \\
217.45233 & 39.07928 & 20.6 \\
217.35072 & 39.16930 & 21.0 \\
\nodata & \nodata & \nodata \\
\enddata
\tablenotetext{1}{Table 5 is published in its entirety in the electronic edition of The 
Astrophysical Journal.  A portion is shown here for guidance regarding its form and content.
This table may also be obtained by request from the authors.}

\end{deluxetable}

\begin{deluxetable}{lccccc}
\tabletypesize{\scriptsize}
\tablewidth{0pt}
\tablecolumns{6}
\tablecaption{Plane Fit Parameters \label{planefits}}
\tablehead{
\colhead{Plane} & \colhead{$a$} & \colhead{$b$} & \colhead{$c$} & \colhead{$d_{GC}$ (kpc)} & \colhead{$d_{sgr}$ (kpc)}
}
\startdata
Sgr North       & -0.20$\pm$0.02   & 0.935$\pm$0.008  & 0.29$\pm$0.01    &  -1.0$\pm$0.3  &  -3.5$\pm$0.7 \\
Sgr South       &  0.03$\pm$0.05   & 0.987$\pm$0.009  & 0.16$\pm$0.05    &  -1.1$\pm$0.5  &   0.7$\pm$1.0 \\
Bif North (N07) & -0.13$\pm$0.05   & 0.980$\pm$0.002  & 0.15$\pm$0.04    &  -0.9$\pm$0.6  &  -1.5$\pm$1.1 \\
Sgr South (K12) & -0.06$\pm$0.07   & 0.97$\pm$0.02    & 0.24$\pm$0.04    &  -1.0$\pm$0.7  &  -1.1$\pm$2.0 \\
Bif South (K12) & -0.19$\pm$0.06   & 0.98$\pm$0.02    & 0.10$\pm$0.04    &  -1.0$\pm$0.6  &  -2.2$\pm$1.9 \\
Majewski (2003) & -0.064$\pm$0.002 & 0.970$\pm$0.008  & 0.233$\pm$0.002  &  0.23$\pm$0.04 &  0.12$\pm$0.07 \\
\enddata
\end{deluxetable}

\begin{deluxetable}{lcccccc}
\tabletypesize{\scriptsize}
\tablewidth{0pt}
\tablecolumns{7}
\tablecaption{Angles Between Planes \label{planeangles}}
\tablehead{
\colhead{Plane} & \colhead{Sgr North ${}^{\circ}$} & \colhead{Sgr South ${}^{\circ}$} & \colhead{Bif North (N07) ${}^{\circ}$} & \colhead{Sgr South (K12) ${}^{\circ}$} & \colhead{Bif South (K12) ${}^{\circ}$} & \colhead{Majewski (2003) ${}^{\circ}$}
}
\startdata
Sgr North       & \nodata   	&	15.6$\pm$0.10 & 9.4$\pm$1.7	& 8.8$\pm$3.8  &  10.9$\pm$2.2 & 8.7$\pm$4.3 \\
Sgr South       & 15.6$\pm$0.10 & 	\nodata		  & 9.6$\pm$0.2 & 7.0$\pm$4.6  &  13.0$\pm$0.7 & 7.0$\pm$3.1 \\
Bif North (N07) & 9.4$\pm$1.7   &	9.6$\pm$0.2   & \nodata  	& 6.5$\pm$3.0  &  3.9$\pm$2.2  & 6.3$\pm$7.6 \\
Sgr South (K12) & 8.8$\pm$3.8   &	7.0$\pm$4.6   & 6.5$\pm$3.0 & \nodata      &  10.4$\pm$0.3 & 0.2$\pm$1.3 \\
Bif South (K12) & 10.9$\pm$2.2  & 	13.0$\pm$0.7  & 3.9$\pm$2.2 & 10.4$\pm$0.3 &  \nodata	   & 10.2$\pm$3.2 \\
Majewski (2003) & 8.7$\pm$4.3   &	7.0$\pm$3.1	  & 6.3$\pm$7.6	& 0.2$\pm$1.3 & 10.2$\pm$3.2 & \nodata	 \\
\enddata
\end{deluxetable}

\begin{deluxetable}{cccc}
\tabletypesize{\scriptsize}
\tablewidth{0pt}
\tablecolumns{3}
\tablecaption{ Spheroid Parameters \label{ spheroidtable }}
\tablehead{
\colhead{Stripe} & \colhead{$q$} & \colhead{$r_0$ (kpc)} }
\startdata
9  & 0.55 $\pm$ 0.17  & 1.84  $\pm$ 2.97 \\
10 & 0.49 $\pm$ 0.13  & 9.82  $\pm$ 5.04 \\
11 & 0.56 $\pm$ 0.13  & 4.65  $\pm$ 2.15 \\
12 & 0.55 $\pm$ 0.10  & 6.94  $\pm$ 2.50 \\
13 & 0.52 $\pm$ 0.11  & 6.08  $\pm$ 4.75 \\
14 & 0.57 $\pm$ 0.11  & 7.15  $\pm$ 4.09 \\
15 & 0.56 $\pm$ 0.09  & 8.59  $\pm$ 4.40 \\
16 & 0.58 $\pm$ 0.13  & 6.22  $\pm$ 4.29 \\
17 & 0.59 $\pm$ 0.13  & 5.37  $\pm$ 5.80 \\
18 & 0.57 $\pm$ 0.12  & 7.32  $\pm$ 4.30 \\
19 & 0.52 $\pm$ 0.12  & 5.99  $\pm$ 5.59 \\
20 & 0.54 $\pm$ 0.11  & 10.14 $\pm$ 4.82 \\
21 & 0.53 $\pm$ 0.14  & 6.45  $\pm$ 3.60 \\
22 & 0.31 $\pm$ 0.16  & 2.88  $\pm$ 4.88 \\
23 & 0.55 $\pm$ 0.07  & 11.55 $\pm$ 3.68 \\
\tableline
79 & 0.34 $\pm$ 0.08 & 25.95 $\pm$ 9.20 \\
82 & 0.46 $\pm$ 0.10 & 19.53 $\pm$ 7.01 \\
86 & 0.63 $\pm$ 0.14 & 16.66 $\pm$ 5.42 \\
\enddata
\end{deluxetable}

\begin{deluxetable}{lcccccccccccccccc}
\tabletypesize{\scriptsize}
\tablewidth{0pt}
\tablecolumns{17}
\tablecaption{ Milkway@home Fits and $N_{\sigma}$}
\tablehead{ \colhead{Stripe} & 
\colhead{$q$} & \colhead{$N_{\sigma}$} & 
\colhead{$r_0$ (kpc)} & \colhead{$N_{\sigma}$} & 
\colhead{$\epsilon$} & \colhead{$N_{\sigma}$} & 
\colhead{$\mu ({}^{\circ})$} & \colhead{$N_{\sigma}$} & 
\colhead{$R$ (kpc)} & \colhead{$N_{\sigma}$} & 
\colhead{$\theta (rad)$} & \colhead{$N_{\sigma}$} & 
\colhead{$\phi (rad)$} & \colhead{$N_{\sigma}$} & 
\colhead{$\sigma$} & \colhead{$N_{\sigma}$} }
\startdata
10 & 0.53 &  0.34 & 8.0  &  0.36 & -1.26 &  0.31 & 230.0 &  1.1  & 36.4 &  1.69   & 1.50 &  0.80 & 3.08 & 0.04  & 5.0 &  2.0 \\
11 & 0.50 &  0.44 & 15.4 &  5.00 & -1.64 &  1.10 & 205.2 &  0.8  & 42.0 &  0.07   & 1.53 &  0.25 & 3.09 & 0.60  & 5.1 &  1.5 \\
12 & 0.55 &  0.02 & 6.77 &  0.07 & -1.34 &  0.16 & 201.6 &  0.1  & 40.6 &  0.01   & 1.31 &  0.02 & 3.13 & 0.10  & 5.5 &  0.2 \\
13 & 0.50 &  0.10 & 6.52 &  0.09 & -1.10 &  0.12 & 197.3 &  0.1  & 40.0 &  0.72   & 1.70 &  1.00 & -2.99 & 0.01 & 4.2 &  0.4 \\
14 & 0.57 &  0.01 & 7.19 &  0.01 & -1.55 &  0.30 & 183.6 &  0.1  & 28.2 &  0.04   & 1.91 &  0.14 & 2.92 & 0.07  & 3.9 &  0.3 \\
15 & 0.56 &  0.00 & 8.64 &  0.01 & -1.80 &  0.17 & 180.1 &  0.01 & 27.6 &  0.03   & 1.91 &  0.01 & 3.00 & 0.00  & 3.3 &  0.1 \\
16 & 0.59 &  0.09 & 5.93 &  0.07 & -1.47 &  0.01 & 169.0 &  0.04 & 28.2 &  0.07   & 2.33 &  0.04 & 2.87 & 0.06  & 2.5 &  0.2 \\
17 & 0.61 &  0.11 & 9.67 &  0.74 & -1.61 &  0.02 & 165.3 &  0.4  & 27.8 &  0.33   & 1.92 &  0.80 & 2.92 & 0.07  & 2.9 &  0.5 \\
18 & 0.56 &  0.08 & 7.32 &  0.00 & -1.97 &  0.00 & 157.3 &  0.03 & 25.2 &  0.02   & 1.97 &  0.18 & 2.82 & 0.18  & 2.3 &  0.2 \\
19 & 0.52 &  0.02 & 5.97 &  0.00 & -1.93 &  0.03 & 151.5 &  0.00 & 23.6 &  0.02   & 2.40 &  0.00 & 3.04 & 0.02  & 1.0 &  0.02 \\
20 & 0.53 &  0.08 & 9.77 &  0.08 & -2.76 &  0.11 & 147.9 &  0.01 & 22.8 &  0.02   & 2.36 &  0.20 & 2.95 & 0.07  & 1.1 &  0.03 \\
21 & 0.53 &  0.03 & 6.65 &  0.06 & -2.01 &  0.29 & 142.1 &  0.03 & 21.4 &  0.02   & 2.62 &  0.02 & 2.84 & 0.08  & 1.1 &  0.1 \\
22 & 0.35 &  0.25 & 3.45 &  0.12 & -2.00 &  0.00 & 135.8 &  0.3  & 17.9 &  0.01   & 2.54 &  0.14 & 1.94 & 0.02  & 1.4 &  0.6 \\
23 & 0.47 &  1.13 & 1.0  &  2.87 & -1.86 &  1.85 & 138.2 &  0.3  & 14.1 &  0.14   & 2.20 &  0.30 & 1.04 & 0.22  & 0.9 &  0.1 \\
\enddata
\end{deluxetable}

\end{document}